# Parallel-disk and cone-and-plate viscometry of a viscoplastic hydrogel with apparent wall slip

Li Quan and Dilhan M. Kalyon*
Stevens Institute of Technology
Hoboken, NJ 07666

*For correspondence: dkalyon@research.stevens.edu

**Abstract**

Hydrogels are widely used in myriad applications including those in the biomedical and personal care fields. It is generally the rheology, i.e., the flow and deformation properties, of hydrogels that define their functionalities. However, the ubiquitous viscoplasticity and associated wall slip behavior of hydrogels handicap the accurate characterization of their rheological material functions. Here parallel-disk and cone-and-plate viscometers were used to characterize the shear viscosity and wall slip behavior of a crosslinked poly(acrylic acid), PAA, hydrogel (carbomer, specifically Carbopol®). It is demonstrated that parallel-disk viscometry, i.e., the steady torsional flow in between two parallel disks, but not the cone-and-plate flow can be used to determine the yield stress and other parameters of viscoplastic constitutive equations and wall slip behavior unambiguously. Parallel-disk viscometry furnishes the yield stress of the hydrogel on the basis of only torque versus the apparent shear rate data. Additional gap-dependent data from parallel-disk viscometry can then be used to characterize the other parameters of the shear viscosity and wall slip behavior of the hydrogel. The accuracies of the parameters of wall slip and Herschel-Bulkley type viscoplastic constitutive equation in representing the flow and deformation behavior of the hydrogel were tested via the comparisons of the calculated (using the lubrication assumption and parallel plate analogy) and experimentally determined velocity distributions of the hydrogel in between two parallel disks and the calculated and experimentally determined torque values. The excellent agreements obtained between the calculated and experimental velocity distributions and experimental and predicted torque values reflect the accuracies of the parameters of shear viscosity and wall slip and indicate that the methodologies demonstrated here provide the means necessary to understand in detail the steady flow and deformation behavior of hydrogels. Such a detailed understanding of the viscoplastic nature and wall slip behavior of hydrogels can then be used to design and develop novel hydrogels with a wider range of applications in the medical and other industrial areas, and for finding optimum conditions for their processing and manufacturing.

**Introduction to gelation and carbomer (Carbopol® hydrogels):**

Both physical and chemical gelation processes are used to generate gels, which exhibit flow and deformation behavior resembling solid elastic bodies and viscous fluids under differing flow conditions. Chemical gelation typically involves a polymerization process whereby the macromolecules are connected (cross-linked) via covalent bonds [Flory, 1953]. Up to a certain degree of conversion the macromolecules are soluble (sol phase) whereas with increasing conversion the macromolecules form a three-dimensional network that spans the entire volume of the sample (gel phase) [Flory, 1953]. As the crosslink density increases during chemical gelation crosslinked polymer clusters are formed and the cluster size increases with increasing degree of crosslinking. When only parts of the polymer molecules crosslink and span the volume, with sol phases in between the macromolecules, a "microgel" is formed [Sperling, 1992]. Thus, microgels can form during conditions when only a few chains are interconnected with each other. Globular proteins can be given as important examples of gelation behavior, i.e., disulfide bonds give rise to the three-dimensional networking of the protein cluster. When the globular clusters are subjected to heat, the intramolecular crosslinks become delocalized giving rise to intermolecular bonds, and high elasticity [Sperling, 1992].

On the other hand, gelation also occurs via physical mechanisms whereby, for example, clusters of particles consisting of crosslinked polymers, start to interact with each other via "dipole-dipole interactions, traces of crystallinity, van der Waals [Budtova *et al.*, 1994; Lu *et al.*, 2008; Trappe *et al.*, 2001; Tanaka *et al.*, 2005; Allain *et al.*, 1995], surface chemistry [Gauckler *et al.*, 1999; Verduin and Dhont, 1995; Grant and Russel, 1993], hydrogen bonding-based complexation [Wang *et al.,* 2017], hydrophobic effects [Cardinaux *et al.*, 2007] and depletion interactions [Bergenholtz *et al.*, 2003; Buzzaccaro *et al.*, 2007; Shah *et al.*, 2003]". Lu *et al.* indicated that during physical gelation particles aggregate to form mesoscopic clusters and networks, as affected by interparticle attractions [Lu *et al.*, 2008]. The onset of the gel state is influenced by three parameters, i.e., first the volume fraction of particles $\phi$, second the $U/k_BT$ ratio, where $U$ is the interparticle attraction strength, and $k_BT$ is the thermal energy (i.e., $k_B$ is the Boltzmann constant and $T$ is the absolute temperature), and third $\xi$, the attractive potential in units of *a*, the particle radius. These three parameters define a three-dimensional state diagram, in which a gelation surface demarcates the boundary between pure liquid like and pure solid like [Lu *et al.*, 2008]. Important attraction mechanisms that drive gelation are generally short range ($\xi$ < 0.1), and include "van der Waals forces, surface chemistry, hydrophobic effects, and some depletion interactions" [Lu *et al.*, 2008]. Gelation requires spinodal decomposition to generate the clusters that span the system [Lu *et al.*, 2008]. During the gelation process when macromolecules go from the sol phase to the gel phase, with various states of microgel formation, the linear viscoelastic properties are good reflections of the structural changes arising from the sol to gel transition [Winter and Chambon, 1986; DeRosa and Winter, 1994; Vural *et al.*, 2010; Wang et al., 2017; Sahin and Kalyon, 2020]. Fully formed gels can be highly elastic.

An interesting gelation agent that is widely used for various industrial applications is carbomer (for example, Carbopol®, which is a tradename of Lubrizol Corporation that is used in our study) which consists of poly(acrylic acid) molecules crosslinked into spherical clusters, i.e., soft particles (Fig. 1). Typically, Carbopol® particles are swollen in water, with a water-rich continuous phase in between the swollen particles [Shafiei *et al.*, 2018].

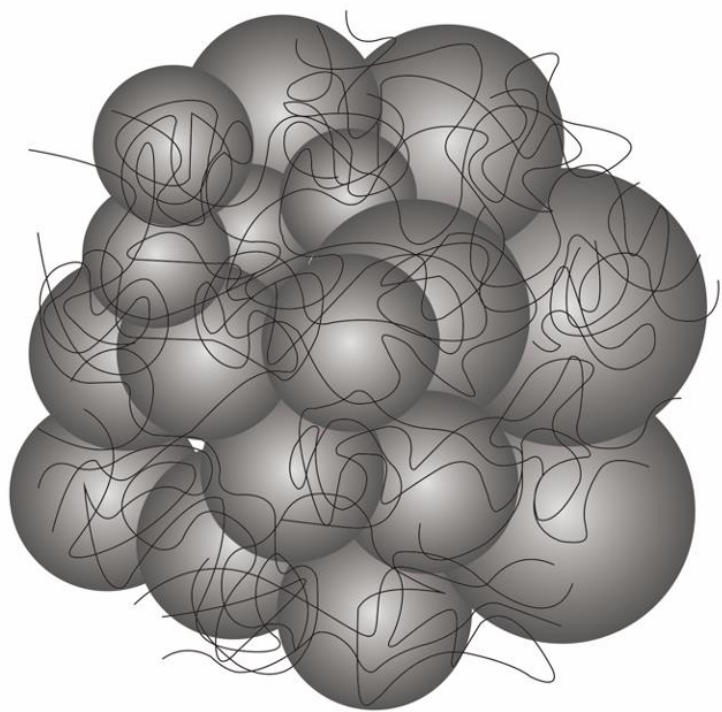

Fig. 1. A schematic representation of the structure of close packed cross-linked and swollen Carbopol gel particles. (Adopted from [Ketz *et al.*, 1988]).

Aqueous dispersions of such crosslinked polymer gels can be prepared over a range of conditions, concentrations and pH. Fig. 2 shows the fluorescence micrographs of Carbopol® hydrogels at various concentrations of Carbopol® [Graziano *et al.*, 2021]. At low concentrations of Carbopol there are no visible interactions and clustering of swollen particles. However, when the concentration of the Carbopol® reaches 0.1% by weight the agglomeration and clustering of the soft particles can be observed [Graziano *et al.*, 2021]. In fact, at this concentration (Fig. 2d) the particle clusters span the length of the sample to generate a microgel environment. The onset of the jamming of the swollen particles give rise to elasticity and gel-like behavior. It should be understood that such network formation is the basis for the development of a yield stress for the hydrogel, which demarcates the boundary between solid-like and fluid-like behavior. The swollen Carbopol® particles were observed to form dendritic aggregates, Fig. 3 [Piau, 2007]. Piau *et al.* have observed that the clustering of the crosslinked particles can span the entire volume above a critical concentration at which a percolated network is developed [Piau, 2007].

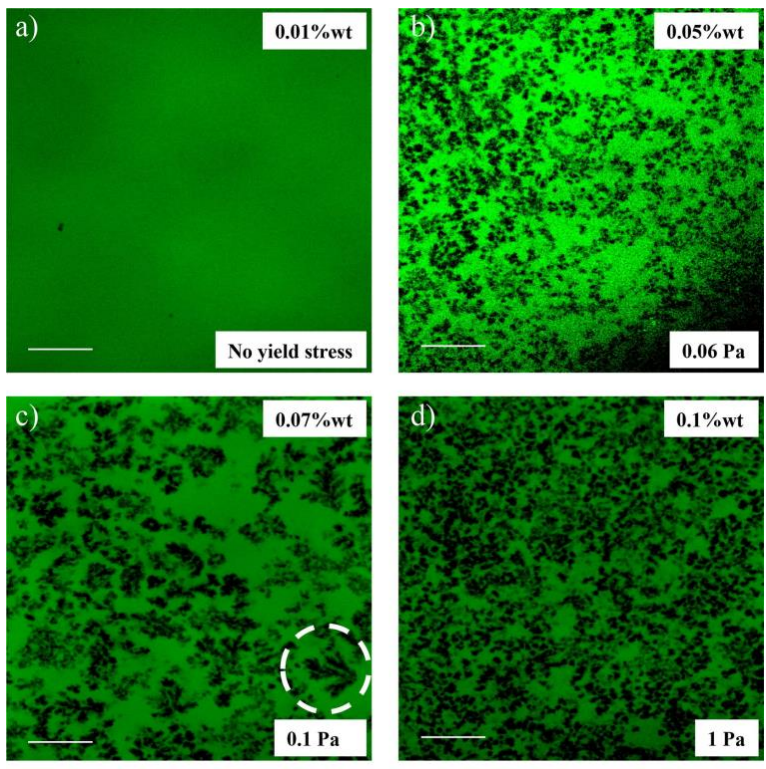

Fig. 2. Carbopol® particles in water phase at various concentrations (0.01 to 0.1wt%) [Graziano *et al.*, 2021]. The scale bar corresponds to 10 µm. Reproduced with permission from Elsevier.

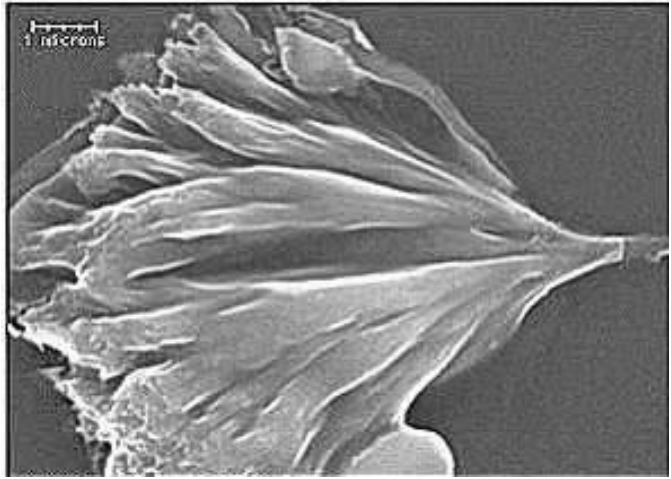

Fig. 3. Carbopol® structure at high magnification after sublimation [Piau, 2007]. Reproduced with permission from Elsevier.

*Viscoplasticity and wall slip of hydrogels*

Carbomer hydrogels (microgels and gels of swollen PAA particles in water) are used in many applications including as thickeners in personal care products [Lochhead, 2007; Rodrigues *et al.*, 2007; Roberts and Barnes, 2001]. The rheological behavior of Carbopol® hydrogels at concentrations around 0.1% and higher have been investigated extensively due to their viscoplastic nature with their flow and deformation behavior affected by the stress field that is acting on the hydrogel during flow [Piau, 2007]. For example, in steady simple shear flow (only one component of the velocity vector prevails and changes only in one other direction) when the absolute value of the shear stress that is applied continuously during a simple shear flow is smaller than the yield stress value of the hydrogel the hydrogel does not deform continuously. Under such conditions a plug flow, enabled by slip at the wall, is observed [Kalyon, 2005; Baek and Kim, 2011]. When the shear stress is greater than the yield stress of the hydrogel there is continuous deformation of the gel accompanied by slip at the wall [Aktas *et al.,* 2014; Baek and Kim, 2011].

Thus, the wall slip and deformation behavior of viscoplastic fluids, including viscoplastic hydrogels, is coupled and need to be investigated concomitantly [Yilmazer and Kalyon, 1991; Kalyon *et al.*, 1993; Kalyon, 1993; Aral and Kalyon, 1994; Yilmazer and Kalyon, 1989; Kalyon *et al.*, 1993; Kalyon, 2003; Tang and Kalyon, 2004; Kalyon, 2005; Kalyon *et al.*, 2006a; Kalyon and Tang, 2007; Kalyon, 2010; Aktas *et al.*, 2014; He *et al.*, 2019]. Generally, the wall slip of complex fluids including suspensions with soft or hard particles and gels occurs via an apparent slip mechanism [Reiner, 1960; Cohen and Metzner, 1985; Jiang *et al.*, 1986; Yilmazer and Kalyon, 1989; Kalyon, 2005], which can also be affected by the presence of a gas phase, i.e., for example, air entrainment [Kalyon *et al.*, 1991a and 1991b; Aral and Kalyon, 1995; Kalyon *et al.*, 1995; Kalyon, 1995]. Such apparent slip layer formation can also be influenced by the migration of particles away from high shear rate regions [Leighton and Acrivos, 1987; Abbott *et al.*, 1991; Nott and Brady, 1994; Koh *et al.*, 1994; Acrivos, 1995; Allende and Kalyon, 2000]. The use of roughened surfaces to eliminate wall slip can lead to the fracture of the viscoplastic fluid [Aral and Kalyon, 1994; Kalyon, 2005a].

There are significant ramifications of apparent wall slip and viscoplastic behavior in complex flows and in processing of various complex fluids [Yilmazer and Kalyon, 1989; Yaras *et al.*, 1994; Lawal and Kalyon, 1992, 1993, 1994a, 1994b, 1997a, 1997b, 1999 and 2000; Lawal *et al.*, 1996; Kalyon *et al.*, 1999; Tang and Kalyon, 2004a, 2004b and 2008; Malik and Kalyon, 2005; Kalyon *et al.*, 2006a, 2006b and 2006c; Kalyon and Malik, 2007 and 2012; Gevgilili *et al.*, 2008; Bonn and Denn, 2009; Kalyon and Malik, 2012; Malik *et al.*, 2014; Aktas *et al.*, 2014]. The wall slip of the polymer phase itself is also observed typically above a critical shear stress. Such slip at the wall of the polymeric binder gives rise to processing difficulties and challenges, including development of flow instabilities that change the nature of the shape of the extrudates emerging from pressure-driven flows, such as shark skin and gross melt fracture and time-dependent pressure [Chen *et al.*, 1993; Gevgilili and Kalyon, 2001; Kalyon *et al.*, 2003; Kalyon *et al.*, 2004; Birinci and Kalyon, 2006; Tang and Kalyon, 2008a and 2008b; Gevgilili *et al.*, 2008].

The flow and deformation behavior of the Carbopol® hydrogels have been investigated earlier in detail via flow through capillary and rectangular slit dies [Pérez-González *et al.*, 2012, Aktas *et al.*, 2014], axial annular flow (flow in between two stationary cylinders as a result of a pressure gradient) [Ortega-Avila *et al.*, 2016], Couette flow (double cylinder one of which is rotating and the other is stationary) [Medina-Bañuelos *et al.*, 2017; Jana *et al.*, 1995; Budtova *et al.*, 1994; Coussot *et al.*, 2009; Divoux *et al.*, 2011 and 2012; Baudez, *et al.*, 2004; Meeker *et al.*, 2004a and 2004b], vane in cup flow [Medina-Bañuelos *et al.*, 2019; Barnes and Carnali, 1990; Estellé *et al.*, 2008; Ovarlez *et al.*, 2011; Nazari *et al.*, 2013; Derakhshandeh *et al.*, 2010]. These viscometric flows have demonstrated the viscoplastic nature of the Carbopol® hydrogels whereby the yield stress value of the hydrogel could be determined unambiguously in conjunction with the wall slip behavior of the hydrogel. In the following an in-depth analysis of the parallel-disk viscometry (steady torsional flow) is carried out to demonstrate how the yield stress of the hydrogel can be determined using parallel-disk viscometry, followed by the characterization of the other parameters of shear viscosity and wall slip and finally the prediction of the torque and velocity distributions in between the two parallel disks, and the comparisons of the predictions of velocity distributions and torques with the experimental values that were available from Medina-Bañuelos *et al.*, 2021.

*Parallel disk viscometry (steady torsional flow)*

Parallel disk viscometry is one of the simplest geometries that can be used for the rheological characterization of complex fluids (steady torsional flow in between two parallel disks) (Fig. 4a) in which the sample is sandwiched in between two disks one of which is rotating at a rotational speed of $\Omega$ and the other is stationary (Fig. 4a). The gap, $H$, in between the two disks is typically significantly smaller than the radius of the disks, $R$, i.e., $H<<R$. The condition, $H<<R$, results in the shear stress component associated with the velocity gradient in the depth direction to be significantly greater than the shear stress component that exists in the radial direction so that the flow can be considered to be a simple shear flow (one component of the velocity vector, $V_\theta$, changing in only one other direction, $z$) and a simple parallel plate analysis of the torsional flow can be carried out [Kalyon, 2021]. It is also possible to use a cone-and-plate fixture whereby the sample is sandwiched in between a cone with a cone angle of $\alpha$ and a disk (Fig. 4b). Typically, the minimum distance of separation between the tip of the truncated cone at its apex and the disk is relatively small, such as 50 μm. As will be shown later the cone and plate geometry is very suitable for the characterization of the rheological behavior of Newtonian and generalized Newtonian fluids that do not exhibit viscoplasticity nor wall slip, since the shear rate and the shear stress are constant within the gap. However, for various complex fluids which exhibit wall slip, like viscoplastic hydrogels, the flow curves are dependent on the radial location, as will be demonstrated later.

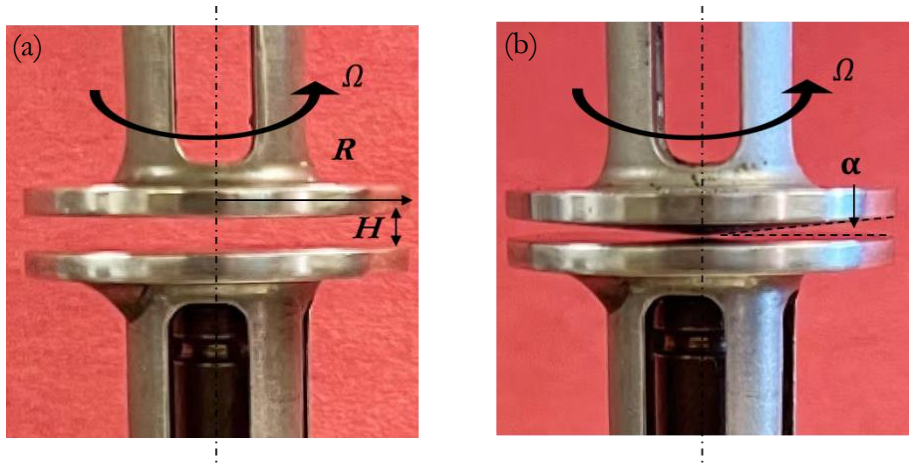

Fig. 4. Steady torsional flow between (a) two parallel disks and between (b) a cone and plate.

In the following first the torque versus the apparent shear rate behavior in steady torsional flow of a Carbopol® hydrogel (at 0.12% by weight Carbopol in water), that was investigated earlier for its Couette [Medina-Bañuelos *et al.*, 2017], and vane in cup [Medina-Bañuelos *et al.*, 2019] flows, are analyzed to generate the parameters of a viscoplastic constitutive equation (Herschel-Bulkley) and the parameters describing the wall slip velocity versus shear stress relationship. It is shown here that the torque versus apparent shear rate data allow in a relatively facile manner, the determination of the yield stress of the fluid, consistent with similar findings from the steady torsional flow of suspensions containing rigid particles [Kalyon, 2021]. Thus, a simple method to determine the value of the yield stress of the hydrogel will be justified and presented. It will be demonstrated that the other parameters of the Herschel-Bulkley fluid constitutive equation can be then determined from the flow curves employing wall slip analysis.

Furthermore, a parallel plate analysis was also used to calculate the velocity distributions and torques as a function of the gap and the rotational speed used during parallel disk viscometry experiments. The velocity distributions in between the disks were calculated and compared with the particle imaging velocimetry data of Medina-Bañuelos and co-workers [Medina-Bañuelos *et al.*, 2021]. The torque values under various flow and geometrical conditions were also calculated. Comparisons between the calculated velocity distributions and torque values, and the experimental velocity distributions and torque data could then be made. Finally, a similar computational analysis based on the parallel plate flow, in conjunction with the lubrication assumption that is valid for relatively small cone angles, was also carried out for cone-and-plate flow (sample sandwiched in between a stationary cone with a relatively small cone angle <4° and a rotating disk). The analysis of the shear stress and shear rate distributions as a function of the radial direction indicated that for the hydrogel neither the shear stress nor the shear rate remain constant within the gap, negating the basic advantage of the cone-and-plate flow. The experimental conveniences of the two flows, i.e., flow between parallel disks and between a cone-and-plate were compared and contrasted.

Most commercial rheometers have the capability to either apply a constant torque value followed by the measurement of the corresponding plate rotational speed or impose a constant rotational speed and measure the corresponding torque. Such rheometers are mistakenly called "constant stress" or constant strain rate rheometers, i.e., "mistakenly", because as will be shown later neither the shear stress nor the strain rate values are known *a priori* in steady torsional flow, unless the fluid is a simple Newtonian fluid, thus what is imposed is either the torque or the rotational speed. Furthermore, the shear stress and the shear rate vary as a function of the radial position. However, it will be shown that in spite of the difficulties this simple shear flow, i.e., parallel-disk viscometry can indeed be analyzed using facile means to generate the parameters of the shear viscosity and wall slip unambiguously. Both constant torque and constant rotational speed rheometers can be used for the methods that are demonstrated here.

**Experimental:**
*Materials*
Carbopol® hydrogels (referred to as "microgels" in some of our earlier publications) were prepared by dissolving different concentrations of Carbopol® 940 in water. A concentration of 0.12% by weight of Carbopol® 940. A fresh batch was obtained from Lubrizol Company. Comparisons of consecutive investigations revealed that freshness is essential for reproducibility of the rheology data. Carbopol® was dissolved in tri-distilled water under continuous stirring [Medina-Bañuelos *et al.*, 2017]. Hollow glass particles (Potters Industries) of 10 μm in size and having a specific gravity of 1.1±0.05 were added into the dispersion at a concentration of 0.03 wt.% to serve as flow tracers. The hydrogel samples prepared this way were used by Medina-Bañuelos *et al.* 2021 to generate the experimentally obtained torques and velocity distributions from parallel-disk viscometry and the torque versus rotational speed data for cone-and-plate viscometry that are used in this manuscript. For the soft particles of the Carbopol® hydrogel with particle radii $a$ (see Fig. 2 and 3 for Carbopol® particles), density, $\rho$, with binder viscosity, $\mu_b$, subject to shear rate $\dot{\gamma}$, with thermal energy, $kT$, the typical particle Reynolds number, *Re*, and the Peclet number, *Pe*, were determined to be $\text{Re}(\dot{\gamma}) = \frac{\rho a^2 \dot{\gamma}}{\mu_b} \ll 1$, $P\text{e}(\dot{\gamma}) = \frac{6\pi\mu_b a^3 \dot{\gamma}}{kT} \gg 1$, suggesting that the steady torsional flow of the Carbopol® hydrogel does take place under the creeping flow regime and that the viscous forces dominate over the colloidal forces over the entire shear rate, $\dot{\gamma}$, range of the experiments.

*Experimental equipment and procedures used by Medina-Bañuelos and co-workers*

The experiments of Medina-Bañuelos *et al.* were carried out using rheo-PIV measurements [Medina-Bañuelos *et al.*, 2021], the rotational rheometer was used in the parallel-disk mode with the sample sandwiched in between two parallel disks. All steady torsional flow measurements were performed at four different gap values, i.e., *H*=0.5 and 0.75, 1.0, and 1.1 mm. The flow curves were obtained by controlling the torque (N-m), in a step-like ramp, and waiting for the attainment of a steady state at each step to record the angular velocity of the upper plate that would correspond to the imposed torque [Medina-Bañuelos *et al.*, 2021]. The PIV data for the steady torsional flow were collected in the *θz* plane as close to the edge of the plates as possible, i.e., at $R_m$=2.35 mm, where $R_m$ is the measuring position of the velocity distributions, as defined as the distance from the edge of the disks. All experiments were carried out at 25±1 °C using fresh samples for each run.

*Yield stress*

As noted earlier when subjected to simple shear flows viscoplastic fluids exhibit solid-like behavior, (plug flow), when the shear stress that is applied is less than a critical value, i.e., the yield stress $\tau_0$, and exhibit a continuous deformation rate when the shear stress applied is above the yield stress. The yield stress develops upon the formation of a three-dimensional network which spans the volume of the viscoplastic fluid. The determination of the yield stress is a challenge and is one of the most misunderstood concepts in the field of rheology.

Various methods based on the linear viscoelastic material functions have been proposed for the estimation of the yield stress $\tau_0$ values of complex fluids [Onogi *et al.*, 1973; Yang *et al.*, 1986; Aral and Kalyon, 1997; Boger, 2013; Kalyon and Aktas, 2014]. For example, the magnitude of the complex shear stress in the plateau region $|\tau_0^*|$, where $|\tau_0^*| = \gamma^0 |G_0^*| = \gamma^0 \left[ \left(G_0^{'}\right)^2 + \left(G_0^{"}\right)^2 \right]^{1/2}$, was taken to represent the yield stress $\tau_0$ [Onogi *et al.*, 1973], $\gamma^0$ is the strain amplitude, $G_0^*$ is the complex shear relaxation modulus, $G_0^{'}$ is the shear storage modulus and $G_0^{"}$ is and shear loss modulus. Another method assumed that $\tau_0$ is equal to the maximum in the plot of the in-phase component of the complex stress, $\tau_0^{'}$, versus the strain amplitude, $\gamma^0$, at low frequencies [Yang *et al.*, 1986].

Flow visualization methods, including magnetic resonance imaging (MRI) [Abbott *et al.*, 1991; Sinton and Chow, 1991; Altobelli *et al.*, 1997; Arola *et al.*, 1997; Raynaud *et al.*, 2002; Coussot *et al.*, 2002; Moraczewski *et al.*, 2005; Bonn *et al.*, 2008], microscopy via a transparent window and a high speed camera [Kalyon *et al.*, 1995], particle tracking velocimetry (PTV) [Holenberg *et al.*, 2012], and particle image velocimetry (PIV) [Holenberg *et al.*, 2012; Pérez-González *et al.*, 2012] offer the capability to visualize conditions under which the viscoplastic fluid exhibits plug flow. When coupled with an experimental method to characterize the shear stress distribution in the flow channel the yield stress can be determined unambiguously.

For example, PIV when used together with capillary rheometry (circular tube, Poiseuille flow) enables the determination of the velocity distributions in the capillary (tubular die) and provides the pressure drop versus flow rate data under fully developed, isothermal, and creeping flow conditions [Pérez-González *et al.*, 2012]. However, the applications of rheometers with flow visualization capabilities are handicapped by the types of fluids that can be characterized, the requisite nature of the rheometer walls, and their spatial resolutions at the wall. For example, MRI requires nonmetallic rheometer walls, and PIV can only be used for the characterization of transparent fluids via transparent walls. Without such flow visualization methods, the characterization of the shear viscosity material function of various complex fluids, including viscoplastic fluids, relies on the characterization of the wall slip velocity versus the wall shear stress [Mooney, 1931] and subsequently the correction of the flow curves (wall shear stress versus the apparent shear rate). These procedures generate the true "wall slip-corrected" shear rates for the characterization of the shear viscosity material function. The wall slip analysis that provides the relationship between the wall slip velocity and the shear stress allows the determination of the shear stress under which plug flow prevails [Yilmazer and Kalyon, 1991], thus making it possible for the yield stress of the viscoplastic fluid to be determined [Kalyon, 2005a; Aktas *et al.*, 2014; Medina-Bañuelos *et al.*, 2017].

In the following steady torsional flow (parallel disk viscometry) is used to determine the yield stress and other viscoplastic flow (Herschel-Bulkley) and wall slip parameters of a Carbopol® hydrogel. The parameters of shear viscosity and wall slip were then tested upon being used to predict the velocity distributions and torques and were compared with the recently published experimental velocity distributions and torque results of Medina-Bañuelos *et al.* [Medina-Bañuelos *et al.*, 2021]. The methods presented here for the characterization of the flow and deformation behavior of viscoplastic fluids have also been tested earlier for a viscoplastic, concentrated suspension of rigid particles [Kalyon, 2021]. Overall, the proposed procedure should significantly simplify the characterization of the yield stress values of viscoplastic hydrogels. The analysis results also provide a better understanding of the flow and deformation behavior of viscoplastic hydrogels in general and their steady torsional flow in particular, for example, by allowing the determination of the shear stress distributions as a function of the radial distance, *r*, and the resulting velocity distributions and torques under various flow conditions in steady torsional flow.

**Background:**

*Viscoplasticity*

Let us start via the formal definition of the flow and deformation behavior of viscoplastic fluids. Viscoplasticity mandates that the flow behavior is binary in nature, i.e., that the deformation rate as represented by the rate of deformation tensor, $\underline{\Delta}$, is zero when the stress magnitude is less than the yield stress, i.e., $1/2(\underline{\tau}:\underline{\tau}) < \tau_0^2$:

$$\underline{\Delta} = 0 \quad \text{for } 1/2(\underline{\tau}:\underline{\tau}) \leq \tau_0^2 \tag{1a}$$

and is finite when the stress magnitude is greater than the yield stress, i.e., for the condition $1/2(\underline{\tau}:\underline{\tau}) > \tau_0^2$:

$$\underline{\tau} = -\eta\left(II_\Delta\right)\underline{\Delta} \quad \text{for } 1/2(\underline{\tau}:\underline{\tau}) > \tau_0^2 \tag{1b}$$

where the shear viscosity, $\eta$, is a function of the second invariant of the rate of deformation tensor, $II_\Delta$, i.e., $\eta\left(II_\Delta\right)$. Equation (1b) is the generalized Newtonian fluid model, which stipulates that the stress tensor is equal to the rate of deformation tensor times the shear viscosity material function [Bird *et al.*, 1987]. It was shown for various viscometric and processing flows that the Herschel-Bulkley equation accurately represents the behavior of various viscoplastic fluids for $1/2(\underline{\tau}:\underline{\tau}) > \tau_0^2$ [Bird *et al.*, 1960; Kalyon *et al.*, 1999; Aktas *et al.*, 2014; Ortega-Avila *et al.*, 2016; Medina-Bañuelos *et al.*, 2017; Tang and Kalyon, 2004 and 2008; Kalyon and Tang, 2007; Kalyon and Aktas, 2014; Medina-Bañuelos *et al.*, 2019], i.e.,

$$\underline{\tau} = -\left(\frac{\tau_0}{\left|\sqrt{1/2(\underline{\Delta}:\underline{\Delta})}\right|} + m\left|\sqrt{1/2(\underline{\Delta}:\underline{\Delta})}\right|^{n-1}\right)\underline{\Delta} \quad \text{for } 1/2(\underline{\tau}:\underline{\tau}) > \tau_0^2 \qquad (1c)$$

The Herschel-Bulkley equation involves three parameters at constant temperature, i.e., the yield stress, $\tau_0$, the consistency index, $m$, and the shear rate sensitivity index, $n$.

For steady torsional flow the Herschel-Bulkley equation becomes:

$$\tau_{z\theta}(r) = \pm\tau_0 - m\left|\frac{dV_\theta}{dz}(r)\right|^{n-1}\left(\frac{dV_\theta}{dz}(r)\right) \quad \text{for} \quad |\tau_{z\theta}(r)| > \tau_0 \qquad (2a)$$

$$\frac{dV_\theta}{dz}(r) = 0 \quad \text{for} \quad |\tau_{z\theta}(r)| \leq \tau_0 \qquad (2b)$$

where $\frac{dV_\theta}{dz}(r)$ is the true shear rate, and $\tau_{z\theta}(r)$ is the shear stress for any radial position, $r$, [Bird *et al.*, 1987; Kalyon, 2005; Aktas *et al.*, 2014]. In Equation (2a) the – sign is used when the shear stress, $\tau_{z\theta}(r)$ is negative. Considering the case of the top disk rotating, so that $\tau_{z\theta}(r) < 0$, Equation (2a) becomes: $\tau_{z\theta}(r) = -\tau_0 - m\left|\frac{dV_\theta}{dz}(r)\right|^{n-1}\left(\frac{dV_\theta}{dz}(r)\right) = -\tau_0 - m\left(\frac{dV_\theta}{dz}(r)\right)^n$.

*Wall slip velocities in steady torsional flow*

Fig. 5a and 5b show the schematics of the velocity distributions for a viscoplastic fluid during steady torsional flow as depicted via the parallel plate flow assumption. Under the conditions of the shear stress, $|\tau_{z\theta}(r)| \leq \tau_0$ and $|\tau_{z\theta}(r)| > \tau_0$, plug flow occurs in between the two walls (Fig. 5a) and continuous deformation rate prevails, respectively.

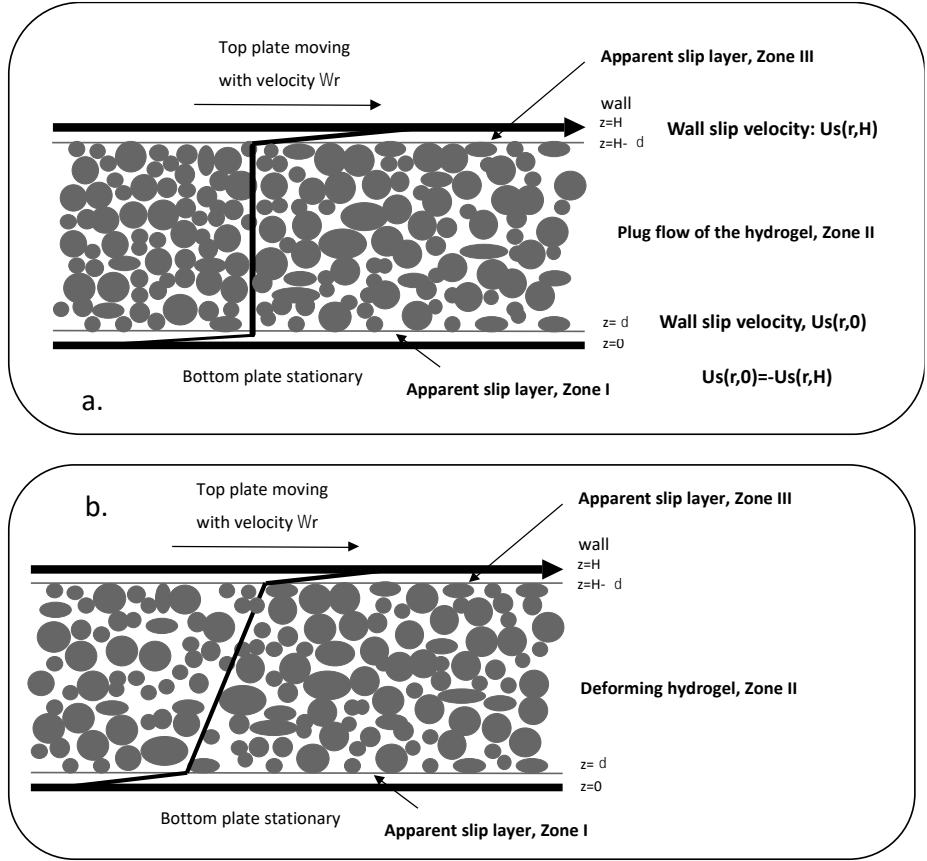

Fig. 5. Schematics of the steady torsional flow of a viscoplastic hydrogel (a) plug flow with apparent slip (b) continuous deformation with apparent slip at the wall.

The apparent wall slip of viscoplastic fluids is also schematically shown in Fig. 5, where apparent slip layers are depicted in an exaggerated manner at both the top and bottom surfaces [Kalyon, 2005]. The wall slip velocity $U_s$ is defined as the difference between the velocity of the fluid at the wall, and the velocity of the wall. Thus, the wall slip velocity is negative for the top disk which is moving, i.e., $U_s(r,H) < 0$, and the wall slip velocity, $U_s(r,0)$, is positive for the bottom disk, which is stationary, i.e., $U_s(r,0) > 0$. The wall slip velocities at the top and bottom disks are related to each other as $U_s(r,H) = -U_s(r,0)$. For the rest of the manuscript "slip velocity" will refer to the absolute values of the wall slip velocities at the top and bottom surfaces to avoid confusion. For similar wall slip behavior at the top and bottom surfaces the slip velocity for plug flow conditions is equal to the wall velocity, $V_w = \Omega R$, over two, i.e., $U_s = \Omega R / 2$ [Kalyon, 2005].

For $|\tau_{z\theta}(r)| \leq \tau_0$, i.e., for plug flow, the slip velocity is only a function of the plate velocity, $\Omega r$ [Kalyon, 2005],

$$U_s(r,0) = \frac{\Omega r}{2} \quad \text{and} \quad U_s(r,H) = -\frac{\Omega r}{2} \tag{3}$$

*Apparent slip flow mechanism*

The wall slip behavior of various viscoplastic fluids, including concentrated suspensions and gels with rigid and soft particles, is subject to the apparent slip mechanism. During the flow of a suspension or gel of rigid or soft particles the particles cannot physically occupy the space adjacent to a wall as efficiently as they can away from the wall. This leads to the formation of a generally relatively thin, but always-present, layer of pure fluid adjacent to the wall, i.e., the "apparent slip layer" or the "Vand layer" [Vand, 1948]. The lower viscosity at the particle-free apparent slip layer gives rise to a higher shear rate at the wall at a given shear stress and hence gives the appearance of wall slip, considering that the slip layer thickness is much smaller than the channel dimension, i.e., apparent wall slip [Rainer, 1960; Cohen and Metzner, 1985; Yilmazer and Kalyon, 1989; Kalyon, 2005].

For suspensions of rigid particles, the estimates of the slip layer thickness over the particle diameter ratio are available [Yilmazer and Kalyon, 1989; Meeker *et al.*, 2004a and 2004b; Kalyon, 2005; Jana *et al.*, 1995; Soltani and Yilmazer, 1998]. Meeker *et al.* have shown that the apparent slip mechanism is also applicable to microgel pastes and concentrated emulsions and have provided methods for the estimation of the apparent slip layer thickness, $\delta$, based on elastohydrodynamic lubrication between squeezed soft particles and shearing surfaces [Meeker *et al.*, 2004a and 2004b]. For viscoplastic microgels the apparent slip mechanism could be integrated into the analysis of various flows including steady torsional, capillary, tangential annular (Couette), axial annular and vane in cup flows [Aktas *et al.*, 2014; Ortega-Avila *et al.*, 2016; Medina-Bañuelos *et al.*, 2017, Medina-Bañuelos *et al.*, 2019].

The relationship between the slip velocity, $U_s(\tau_{z\theta}(r))$, and the shear stress, $\tau_{z\theta}(r)$, for apparent wall slip occurring in steady torsional flow becomes the following for Vand layers, the shear viscosity of which can be represented by a power law equation represented with a consistency index, $m_b$, and a power law index of $n_b$, i.e., $\tau_{z\theta}(r) = -m_b \left|\frac{dV_\theta}{dz}(r)\right|^{n_b-1}\left(\frac{dV_\theta}{dz}(r)\right) = -m_b\left(\frac{dV_\theta}{dz}(r)\right)^{n_b}$, as [Kalyon, 2005]:

$$U_s(r,0) = \frac{\delta}{m_b^{1/n_b}}\left(-\tau_{z\theta}(r)\right)^{1/n_b} \text{ and } U_s(r,H) = -\frac{\delta}{m_b^{1/n_b}}\left(-\tau_{z\theta}(r)\right)^{1/n_b} \quad (4)$$

It is important to note that the relationship between the slip velocity and wall shear stress remains the same (Eq. 4), provided that that apparent slip layer thickness, $\delta$, remains a constant regardless of the flow conditions. Thus, with $\delta$ constant the wall slip behavior would remain the same whether the flow is occurring in the plug flow region or in the continuous deformation region. For suspensions of rigid, low aspect ratio and non-colloidal particles in the volume fraction of solids, range of 0.17 to 0.94 compilation of apparent slip layer thickness data over a wide range of concentrations have indicated that the apparent slip layer can be related to the harmonic mean particle diameter and the ratio of the volume fraction of solids over their maximum packing fraction, i.e., $\frac{\phi}{\phi_m}$, and can be determined from: $\frac{\delta}{D_p} = \left(1 - \frac{\phi}{\phi_m}\right)$ [Kalyon, 2005; Ballesta et al., 2008]. For pressure-driven flows the behavior under the plug flow conditions can be complicated and the apparent slip layer can be a function of the flow rate [Yaras et al., 1994]. Such dependence on the flow conditions can be a consequence of the binder itself exhibiting wall slip, which typically occurs at shear stresses that are above a critical wall shear stress [Giesekus and Langer, 1977; El Kissi and Piau, 1990; Brunn and Vorwerk, 1993; Tang and Kalyon, 2008a and 2008b]. However, such complications are not observed for Newtonian binders [Yilmazer and Kalyon, 1989; Jana et al., 1995; Kalyon, 2005; He et al., 2019]. It should be noted that in the following additional light will be shed onto the nature of the apparent wall slip mechanism for carbomer hydrogels.

**Results and Discussion:**

The torque, $\Im$, versus the apparent shear rate data ($\dot{\gamma}_{aR} = \Omega R/H$) from parallel-disk viscometry are shown in Fig. 6a for three gaps of 0.5, 0.75 and 1 mm [Medina-Bañuelos et al., 2021]. The data were best fitted to determine the variation of the slope, $\frac{d\ln\mathfrak{I}}{d\ln(\Omega R/H)}$, for the entire apparent shear rate range of 0.5 to 100 s$^{-1}$. There are two distinct slopes, the first is valid for all gaps for torques less than $7*10^{-4}$ to $9*10^{-4}$ N-m and the second slope prevails above this range. Thus, for all gaps the slope $\frac{d\ln\mathfrak{I}}{d\ln(\Omega R/H)}$ changes at the critical torque, $\mathfrak{I}_c$ (Fig. 6b). For $\mathfrak{I} \geq \mathfrak{I}_c$, the slope of the torque, $\mathfrak{I}(H, \Omega)$, versus the apparent shear rate at the edge, $\Omega R/H$, i.e., $\frac{d\ln\mathfrak{I}}{d\ln(\Omega R/H)}=0.50$ for $H=1$ mm, $\frac{d\ln\mathfrak{I}}{d\ln(\Omega R/H)}=0.53$ for $H=0.75$ mm, and $\frac{d\ln\mathfrak{I}}{d\ln(\Omega R/H)}=0.60$ for $H=0.5$ mm. What does this transition in torque represent?

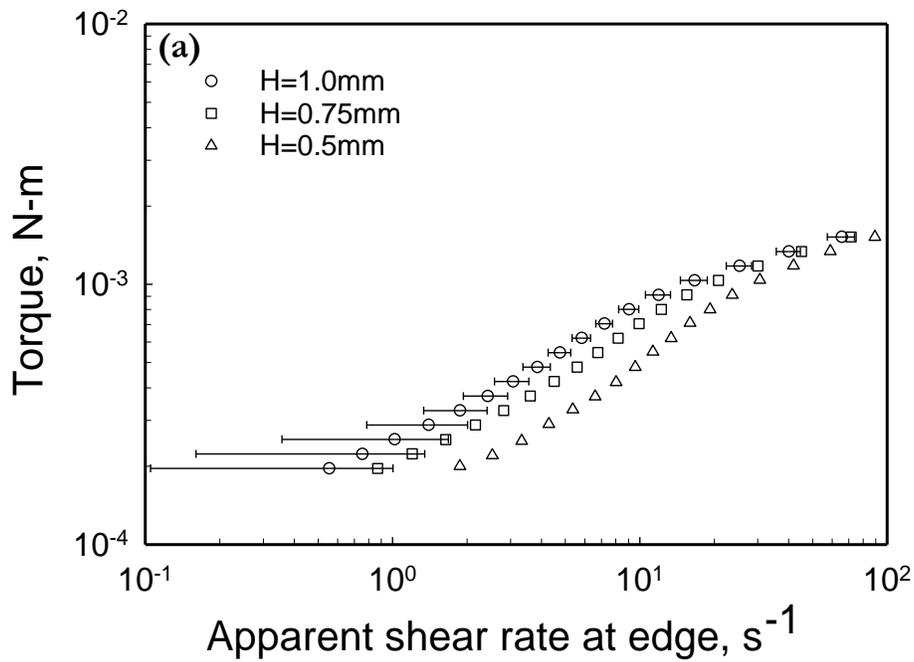

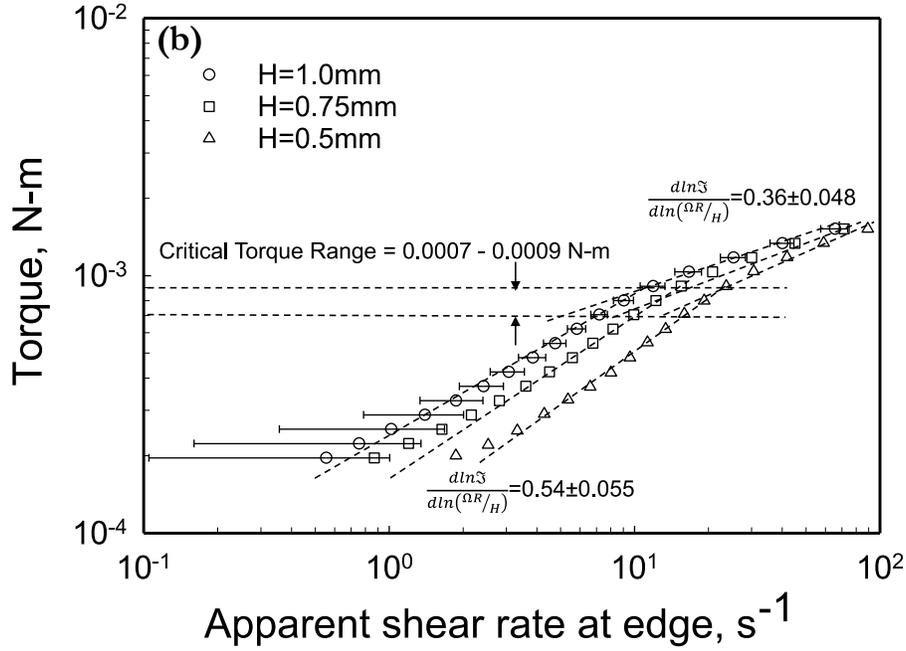

Fig. 6. (a) Steady torque, $\Im$, versus the apparent shear rate at the edge for three gaps. (b) The critical torque corresponds to the yield condition from which the yield stress can be determined.

The torque, $\Im$, that is necessary to rotate the upper disk at a given apparent shear rate at the edge, $\dot{\gamma}_{aR} = \Omega R / H$, is given by:

$$\Im = 2\pi \int_0^R \left(-\tau_{z\theta}(r)\right) r^2 dr \tag{5a}$$

Upon a change of variable of integration to:

$$\frac{\dot{\gamma}_{aR}^3 \Im}{2\pi R^3} = \int_0^{\dot{\gamma}_{aR}} \left(-\tau_{z\theta}(r)\right) \dot{\gamma}_{ar}^2 d\dot{\gamma}_{ar} \tag{5b}$$

and differentiation with respect to the apparent shear rate at the edge, $\dot{\gamma}_{aR} = \Omega R / H$, and utilizing the Leibniz rule of integration, one obtains [Bird *et al.*, 1987] the following relationship between the torque, versus the shear stress at the edge, *R*:

$$-\tau_{z\theta}(R) = \frac{\Im}{2\pi R^3}\left(3 + \frac{d \ln \Im}{d \ln (\Omega R / H)}\right) \tag{6}$$

For the apparent slip mechanism and using a parallel plate analogy, the shear stress $\tau_{z\theta}(r)$ at any radial position, $r$, in steady torsional flow can be determined for the case of the top surface moving using [Kalyon, 2005]:

$$\left[\frac{-(\tau_{z\theta}(r)+\tau_0)}{m}\right]^{1/n}\left(1-\frac{2\delta}{H}\right)+\frac{2\delta}{H}\left(\frac{-\tau_{z\theta}(r)}{m_b}\right)^{1/n_b}=\frac{\Omega r}{H} \quad \text{for } -\tau_{z\theta}(r) > \tau_0 \quad (7a)$$

$$\frac{2\delta}{H}\left(\frac{-\tau_{z\theta}(r)}{m_b}\right)^{1/n_b}=\frac{\Omega r}{H} \quad \text{for } -\tau_{z\theta}(r) \leq \tau_0 \quad (7b)$$

Equations (7a) and (7b) indicate that the relationship between the shear stress, $|\tau_{z\theta}(r)|$ and the apparent shear rate expected for the pure plug flow, i.e., for $|\tau_{z\theta}(r)| \leq \tau_0$, would be different than the one that prevails under shear stresses for which $|\tau_{z\theta}(r)| > \tau_0$. How would this manifest itself for the torque, $\Im$ versus the apparent shear rate at the edge, $\Omega R/H$, behavior and how different would the slope $\frac{d \ln \Im}{d \ln(\Omega R/H)}$ be for the continuous deformation rate region, i.e., $|\tau_{z\theta}(r)| > \tau_0$ in comparison to the plug flow region, i.e., $|\tau_{z\theta}(r)| \leq \tau_0$?

For the apparent wall slip mechanism the torque values for pure plug flow, i.e., $|\tau_{z\theta}(R)| \leq \tau_0$, can be determined as a function of the apparent shear rate at the edge of the disks, $\dot{\gamma}_{aR} = \Omega R/H$, as the following for a binder of viscoplastic fluids with a power-law type shear viscosity with the consistency index, $m_b$ and a power-law index, $n_b$ (for constant apparent slip layer thickness, $\delta$):

$$\Im(r_0 > R) = \frac{2\pi m_b R^3}{(3+n_b)}\left(\frac{\Omega R}{2\delta}\right)^{n_b} = \frac{2\pi m_b R^3 H^{n_b}}{(3+n_b)(2\delta)^{n_b}}(\dot{\gamma}_{aR})^{n_b} \quad (8a)$$

Thus, for a non-Newtonian binder that constitutes the apparent slip layer with constant $\delta$, and with shear viscosity represented by a power-law equation, the slope $\dfrac{d \ln \Im}{d \ln(\Omega R/H)}$ would be equal to the power law index of the binder, $n_b$.

On the other hand, for a Newtonian binder with viscosity, $\mu_b$.

$$\Im(r_0 > R) = \frac{\pi \mu_b R^4 \Omega}{4\delta} = \frac{\pi \mu_b R^3 H}{4\delta}\dot{\gamma}_{aR} \qquad (8b)$$

and hence $\dfrac{d \ln \Im}{d \ln(\Omega R/H)} = 1$.

What happens if the shear stress, $|\tau_{z\theta}(r)|$ exceeds the yield stress at a radial position $r_0$ during steady torsional flow, so that part of the viscoplastic fluid in between the two parallel disks is undergoing plug flow, i.e., for $r \leq r_0$, and part of the fluid is undergoing continuous deformation for $r > r_0$, with a transition at $r = r_0$, i.e., $|\tau_{z\theta}(r_0)| = \tau_0$? For cases where there is plug flow and continuous deformation occurring simultaneously the torque, $\Im$, can be determined as [Kalyon, 2021]:

$$\Im = 2\pi \int_0^R (-\tau_{z\theta}(r))r^2 dr = 2\pi \left[ \int_0^{r_0} (-\tau_{z\theta}(r))r^2 dr + \int_{r_0}^R (-\tau_{z\theta}(r))r^2 dr \right] \qquad (9a)$$

where $r_0$ is the radial location at which $|\tau_{z\theta}(r_0)| = \tau_0$. The first term on the right is thus the contribution of the flow in the plug flow zone to the torque and is equal to:

$$2\pi \int_0^{r_0} (-\tau_{z\theta}(r))r^2 dr = \frac{2\pi m_b r_0^{3+n_b}}{(3+n_b)} \left( \frac{\Omega}{2\delta} \right)^{n_b}. \qquad (9b)$$

An analytical expression for the second integral on the right of Equation (9a) can be substituted if in the continuous deformation region of steady torsional flow, i.e., for $|\tau_{z\theta}(r)| > \tau_0$, the wall slip contribution to shear stress is assumed to be negligible in comparison to the contribution of bulk deformation of the suspension to shear stress:

$$\int_{r_0}^{R}(-\tau_{z\theta}(r))r^2 dr = \frac{\tau_0}{3}(R^3 - r_0^3) + \frac{m}{(3+n)}\left(\frac{\Omega}{H}\right)^n (R^{3+n} - r_0^{3+n}) \qquad (9c)$$

so that the torque, $\Im$, for the condition of negligible slip contribution in the continuous deformation region becomes [Kalyon, 2021]:

$$\Im = 2\pi \int_{0}^{R}(-\tau_{z\theta}(r))r^2 dr = 2\pi\left[\frac{m_b r_0^{3+n_b}}{(3+n_b)}\left(\frac{\Omega}{2\delta}\right)^{n_b} + \frac{\tau_0}{3}(R^3 - r_0^3) + \frac{m}{(3+n)}\left(\frac{\Omega}{H}\right)^n (R^{3+n} - r_0^{3+n})\right]$$

(9d)

The third term on the right side of Equation (9d) dominates for $R \gg r_0$ so that $\frac{d\ln\Im}{d\ln(\Omega R/H)} \approx n$ ($n$ being the shear rate sensitivity index of the Herschel-Bulkley fluid (Eq. 1 and Eq. 2)).

Thus, in the steady torsional flow of the hydrogel subject to apparent wall slip a slope change in $\frac{d\ln\Im}{d\ln(\Omega R/H)}$ from the power law index of the binder, $n_b$, to a value approaching the shear rate sensitivity index, $n$, of the Herschel-Bulkley fluid. Typically, $n < n_b$, considering that when a binder is mixed with soft (as in the hydrogel of this study) or rigid particles [He et al., 2019] the resulting suspension is generally pseudoplastic in nature, i.e., $n<1$. There are exceptions to this for dilatant suspensions for which $n>1$. Such dilatant suspensions typically incorporate low-aspect particles with a narrow size range [Yilmazer and Kalyon, 1989 and 1991]. Regardless of the nature of the rheological behavior of the viscoplastic fluid versus the rheological behavior of the binder the change in slope, $\frac{d\ln\Im}{d\ln(\Omega R/H)}$, reflects the transition from pure plug flow to a flow with both plug flow for $r \leq r_0$ and continuous deformation for $r_0 < r \leq R$.

Therefore, the change in the slope, $\dfrac{d\ln\Im}{d\ln(\Omega R/H)}$, is expected to occur when the shear stress at the edge becomes equal to the yield stress, i.e., $|\tau_{z\theta}(R)|=\tau_0$. Thus, this step change in the slope $\dfrac{d\ln\Im}{d\ln(\Omega R/H)}$ serves as the basis for the determination of the yield stress, $\tau_0$, values of viscoplastic fluids using steady torsional flow [Kalyon, 2021]. It is sufficient for the determination of the yield stress to collect the torque, $\Im$, versus rotational speed, $\Omega$, data at a single gap, $H$.

For the hydrogel what is the critical shear stress range at the edge, $|\tau_{z\theta}(R)|_c$, that corresponds to the critical torque range of $0.0007 \leq \Im_c \leq 0.0009$ N-m? Applying Equation (6), i.e., $|\tau_{z\theta}(R)|=\dfrac{\Im}{2\pi R^3}\left(3+\dfrac{d\ln\Im}{d\ln(\Omega R/H)}\right)$ for the critical condition, i.e., $|\tau_{z\theta}(R)|_c=\dfrac{\Im_c}{2\pi R^3}\left(3+\dfrac{d\ln\Im}{d\ln(\Omega R/H)}\right)$ with an average for the three gaps value of $\dfrac{d\ln\Im}{d\ln(\Omega R/H)}$ =0.54 for $\Im<\Im_c$ (Fig. 6b) the critical shear stress range at the edge, $|\tau_{z\theta}(R)|_c$ is determined to be 24-30 Pa (Fig. 7). Thus, the yield stress, $\tau_0$, of the hydrogel is about 27 Pa, which is exactly what was determined as the yield stress of this hydrogel from previous investigations using Couette flow [Medina-Bañuelos *et al.*, 2017] and vane in cup flow [Medina-Bañuelos *et al.*, 2019]. The new methodology that is applied here for the determination of the yield stress of a viscoplastic fluid using steady torsional flow was tested earlier for a concentrated suspension of rigid particles mixed with a poly(dimethyl siloxane) binder [Kalyon, 2021]. In that investigation the determined yield stress value using the torque versus apparent shear rate data from steady torsional flow was found to be similar to the yield stress values of the concentrated suspension obtained using wall slip analysis, as well as a straight-line marker method [He *et al.*, 2019].

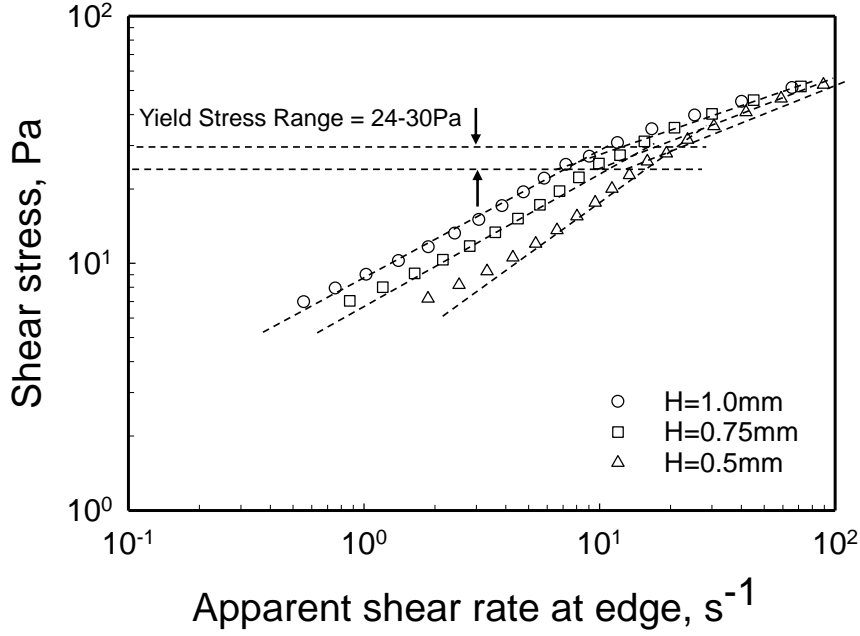

Fig. 7. Shear stress at the edge, $|\tau_{z\theta}(R)|$, versus the apparent shear rate at the edge, $\dot{\gamma}_{aR} = \Omega R/H$, for three gaps, $H$=1, 0.75 and 0.5 mm the yield stress range is indicated and corresponds to the critical torque range.

*Apparent slip analysis*

It was indicated earlier that for the conditions of the apparent slip layer thickness, $\delta$, or the shear viscosity behavior of the fluid comprising the apparent slip layer thickness (for a power-law fluid consistency index, $m_b$, and power-law index, $n_b$) remaining the same over the rotational speed, $\Omega$, range imposed during plug flow of the hydrogel Equation (8a) would be valid:

$\Im(r_0 > R) = \dfrac{2\pi m_b R^3}{(3+n_b)}\left(\dfrac{\Omega R}{2\delta}\right)^{n_b} = \dfrac{2\pi m_b R^3 H^{n_b}}{(3+n_b)(2\delta)^{n_b}}\left(\dot{\gamma}_{aR}\right)^{n_b}$ . This points out that the torque would remain independent of the gap, $H$, used in the steady torsional flow, regardless of whether the binder fluid is Newtonian or non-Newtonian (note that $H\dot{\gamma}_{aR} = \Omega R$).

However, as shown in Fig. 8 there is dependence of the torque on the gap in the plug flow region indicating that either the slip layer thickness is changing or that the rheological behavior of the fluid constituting the apparent slip layer is changing as the flow conditions are altered. Let us analyze the slip behavior in plug flow further.

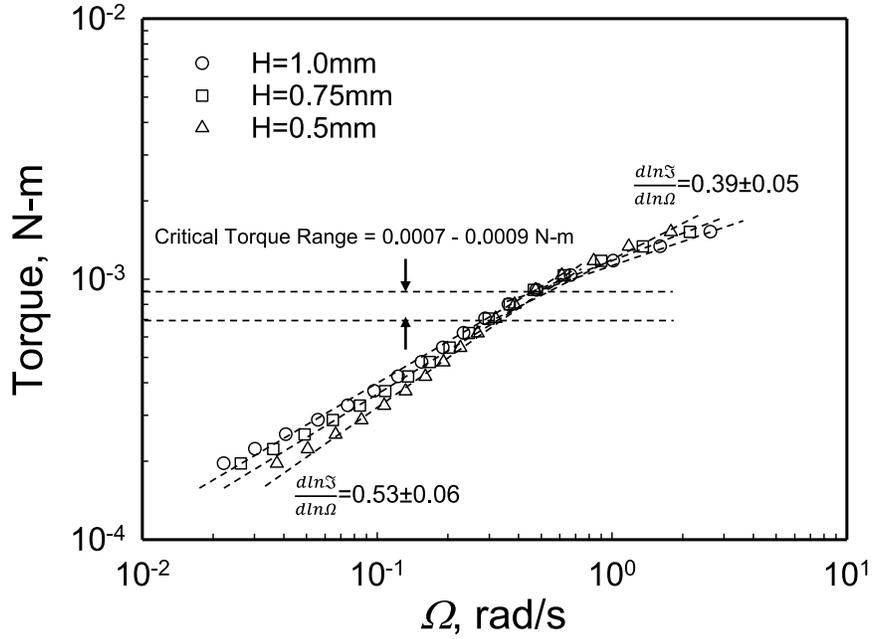

Fig. 8. Steady torque, $\Im$, versus the rotational speed, $\Omega$, at the edge for three gaps.

In general, the wall slip velocity versus the shear stress behavior of complex fluids, including viscoplastic fluids, can be analyzed via systematic changes in the surface to volume ratio of the viscometer, i.e., by changing the gap, $H$ [Yoshimura and Prud'homme, 1988; Yilmazer and Kalyon, 1989] akin to the method suggested by Mooney for flow through circular tubes [Mooney, 1931]:

$$\frac{\Omega r}{H} = \frac{U_s(r,0)}{H} - \frac{U_s(r,H)}{H} + \frac{dV_\theta}{dz}(r) \qquad (10a)$$

$$\frac{\Omega R}{H} = \frac{U_s(R,0)}{H} - \frac{U_s(R,H)}{H} + \frac{dV_\theta}{dz}(R) \qquad (10b)$$

where $\frac{\Omega r}{H}$ is the apparent shear rate, $\dot{\gamma}_{ar}$, at the radial position, $r$, $\frac{\Omega R}{H}$ is the apparent shear rate at the edge, $\dot{\gamma}_{aR}$, and $\frac{dV_\theta}{dz}(R)$ is the true shear rate, $\dot{\gamma}(R)$, imposed on the fluid at $r=R$, i.e., corresponding to the shear stress at the edge, $\tau_{z\theta}(R)$.

The slopes of the apparent shear rate with respect to $1/H$ at constant shear stress provide the absolute value of the wall slip velocity at the given shear stress so that one can obtain the slip velocity versus the shear stress behavior.

$$\left.\frac{\partial\left(\Omega R/H\right)}{\partial\left(1/H\right)}\right|_{\tau_{z\theta}} = 2U_s\left(\tau_{z\theta}(R)\right) \tag{10c}$$

Equation (10c) suggests that if plots of apparent shear rate versus reciprocal gap are drawn at constant shear stress at the edge, the slopes would be equal to $2U_s\left(\tau_{z\theta}(R)\right)$ and extrapolated intercepts would be equal to the true shear rate at the edge. Yilmazer and Kalyon [Yilmazer and Kalyon, 1989] have used more than two gaps and thus utilized Equation (11) whereas Yoshimura and Prud'homme have used only two gaps in their analysis [Yoshimura and Prud'homme, 1988] so that:

$$U_s\left(\tau_{z\theta}(R)\right) = \pm R \frac{\left[\dfrac{\Omega_1\left(\tau_{z\theta}(R)\right)}{H_1} - \dfrac{\Omega_2\left(\tau_{z\theta}(R)\right)}{H_2}\right]}{2\left(\dfrac{1}{H_1} - \dfrac{1}{H_2}\right)} \tag{11}$$

where $\Omega_1\left(\tau_{z\theta}(R)\right)$ and $\Omega_2\left(\tau_{z\theta}(R)\right)$ are the rotational speeds for the two gaps, $H_1$ and $H_2$, at the same shear stress, $\tau_{z\theta}(R)$.

*Plug flow*

Starting with Fig. 7 the application of the analysis contained in Equation (10), i.e., for each gap, $H$, the apparent shear rate versus $1/H$ data were used at various shear stress values to determine the slopes which are equal to $2U_s$ to determine the relationship between slip velocity and shear stress. Fig. 9 shows the slip velocity versus the shear stress behavior of the hydrogel determined in the plug flow region, i.e., $|\tau_{z\theta}(r)| \leq \tau_0$. The y-intercept in Equation (10b) represents the true shear rate of the hydrogel. For plug flow the y-intercept should be zero for data collected at all three gaps, indicating that plug flow prevails and the true shear rate is equal to zero. This expected behavior is indeed observed. As would be expected from the data shown in Fig. 8 the slip velocity values obtained at different gaps, although are close to each other, suggest some degree of dependence of the slip velocity values to the conditions generated at the different gaps that were used.

Meeker *et al.* have analyzed the formation of the apparent slip layer for gels with soft particles [Meeker *et al.*, 2004a, 2004b; Aktas *et al.*, 2014]. For the plug flow formation in steady torsional flow, Meeker *et al.* have determined based on Reynolds lubrication equation that the apparent slip layer, $\delta$, can be given as:

$$\delta = \left(\frac{\mu_w U_s R_p}{G_p}\right)^{1/2} \tag{12a}$$

where the Carbopol® microgel with a Newtonian binder (water), with shear viscosity $\mu_w$ consists of closely packed swollen soft particles with modulus of elasticity of $G_p$ and radius, $R_p$ [Meeker *et al.*, 2004a, 2004b]. $\tau_{z\theta}$ can be given as:

$$\tau_{z\theta} = \frac{\mu_w U_s}{\delta} = \frac{\mu_w U_s}{\left(\frac{\mu_w U_s R_p}{G_p}\right)^{1/2}} = \left(\frac{\mu_w U_s G_p}{R_p}\right)^{1/2} \tag{12b}$$

and hence:

$$\frac{\delta}{R_p} = \left(\frac{1}{G_p}\right)\tau_{z\theta} \tag{12c}$$

Aktas *et al.* [Aktas *et al.*, 2014] have determined that for Carbopol® hydrogels the ratio of $R_p$ over $G_p$ is a constant for the plug flow region, i.e., the apparent slip layer thickness varies linearly with the shear stress. A corollary of this finding is that the apparent slip velocity $U_s$ would vary with the square of the shear stress, i.e., $U_s = \tau_{z\theta}^2$ in the plug flow region. As shown in Fig. 9 the exponent is in the range of 1.50 to 1.65 depending on the gap and the method used, and is thus smaller than 2, indicating that there is another mechanism at play.

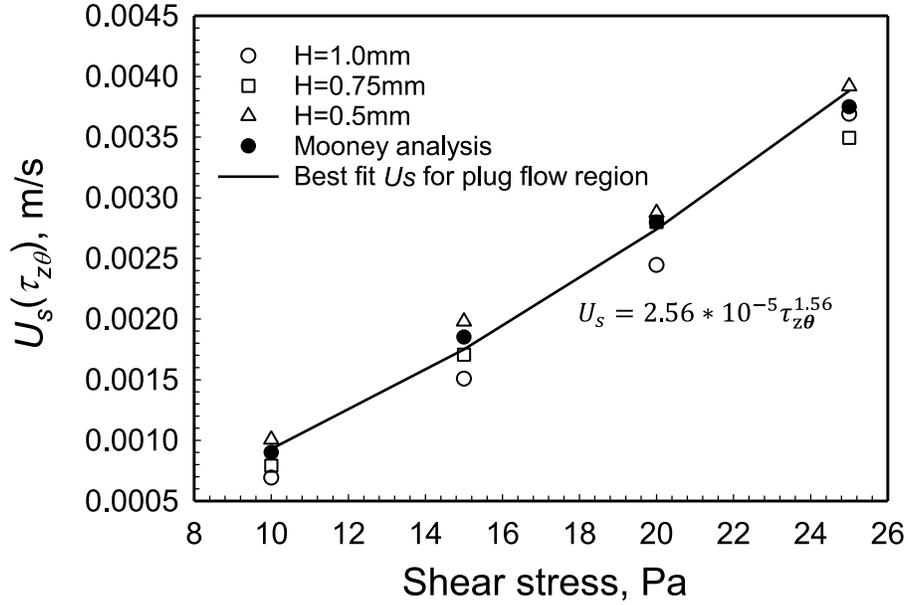

Fig. 9. Wall slip velocity, $U_s$, versus shear stress at the edge, $|\tau_{z\theta}(R)|$, from the data collected at three gaps in the plug flow region.

As indicated earlier in Equation (4) the relationship between the slip velocity, $U_s$, and the shear stress, $\tau_{z\theta}(R)$, for steady torsional flow is equal to $U_s(R) = \pm \beta(-\tau_{z\theta}(R))^{s_b}$ [Kalyon, 2005], with $\pm$ necessary to accommodate the changing sign of the slip velocity at the stationary and moving walls, $\beta$ is the slip coefficient and the reciprocal power law index, $s_b = 1/n_b$ of the fluid that constitutes the apparent slip layer. For the gap dependency to be present in the above analysis either the apparent slip layer thickness, $\delta$, or the shear viscosity of the fluid constituting the apparent slip layer should change under different flow conditions- although all lead to plug flow of the hydrogel. The slope $s_b$ gives a hint as to what is happening. Considering that the binder of the gel is Newtonian water, and therefore $n_b=1$ and hence $1/n_b= s_b=1$. However, as seen in Fig. 9 the value of the slope, $s_b$, for plug flow is in the range of 1.5 to 1.65, and thus $n_b$ for the apparent slip layer in plug flow region is 0.60 to 0.67. This finding suggests that the fluid that constitutes the apparent slip layer for the plug flow region is non-Newtonian. What could impart a non-Newtonian character to the apparent slip layer, if the major constituent is water?

It is reasonable to assume that the soft, cross-linked, spherical PAA particles with dangling chains attached to their surfaces cannot come and pack efficiently at the wall as they can away from the wall. However, the free end of the PAA chains can penetrate into the apparent slip layer under the mild shear stress and shear rate conditions of plug flow, giving rise to a PAA solution at the apparent slip layer thickness. Thus, our hypothesis is that the dangling, poly(acrylic acid), PAA, macromolecules of the Carbopol® hydrogel, that are fixed to the cross-linked particles on one end, are able to rotate and orient freely on the other end. The chains would have some motion and orientation capabilities to penetrate into the apparent slip layer under plug flow conditions as depicted schematically in Fig. 10a.

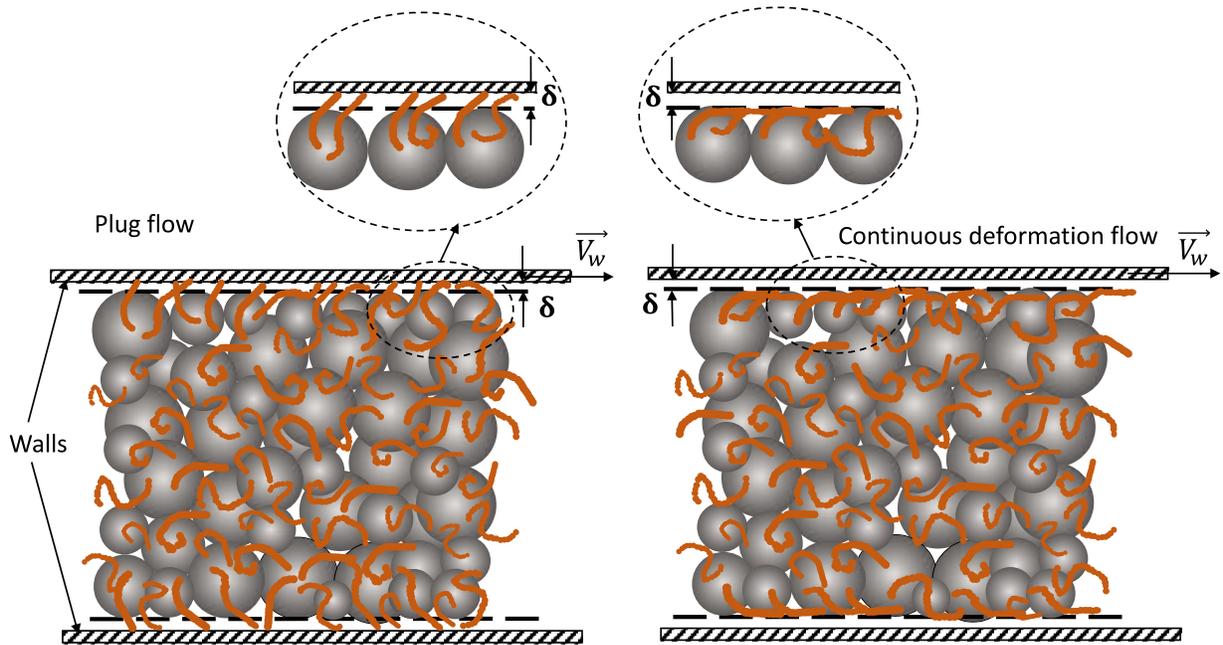

Fig. 10. Hypothetical explanation of the differences in the shear viscosity of the fluid at the apparent slip layer, i.e., (a) penetration of the PAA chains into the slip layer for the plug flow region and (b) apparent slip layer free of particles and PAA chains for the continuous deformation region.

On the other hand, for the continuous deformation region the slope $s_b \approx 1$ and hence the power law index of the fluid constituting the apparent slip layer, $n_b \approx 1$, characteristic of a Newtonian fluid. This is shown in Fig. 11 where it is indicated that for the data used involving the gap $H= 1$ and 1.1 mm, the relationship between the slip velocity and the shear stress is: $U_s = 1.57*10^{-4} \tau_{z\theta}(R)^{0.98}$.

It can be hypothesized that the higher shear stress and the shear rates found in the continuous deformation region of the steady torsional flow orients the macromolecules that are anchored to the soft particle surfaces along the streamlines of the flow field (which are parallel to the wall velocity). This generates an apparent slip layer that is free of particles, as well as free from the presence of dangling PAA macromolecules (Fig. 10b). Thus, only water constitutes the apparent slip layer for the continuous deformation region. Following up on this hypothesis the apparent slip layer thickness, $\delta$, for the continuous deformation region, comprised of water, can be determined by using: $U_s = \beta * |\tau_{z\theta}(R)| = \frac{\delta}{\mu_w} * |\tau_{z\theta}(R)|$, i.e., $\delta = \beta * \mu_w = 0.16 \mu m$. This thickness determined for the continuous deformation region is a reasonable estimate of the apparent slip layer thickness since the diameter of the soft crosslinked swollen PAA particles are estimated to be in the 2 to 3 μm range.

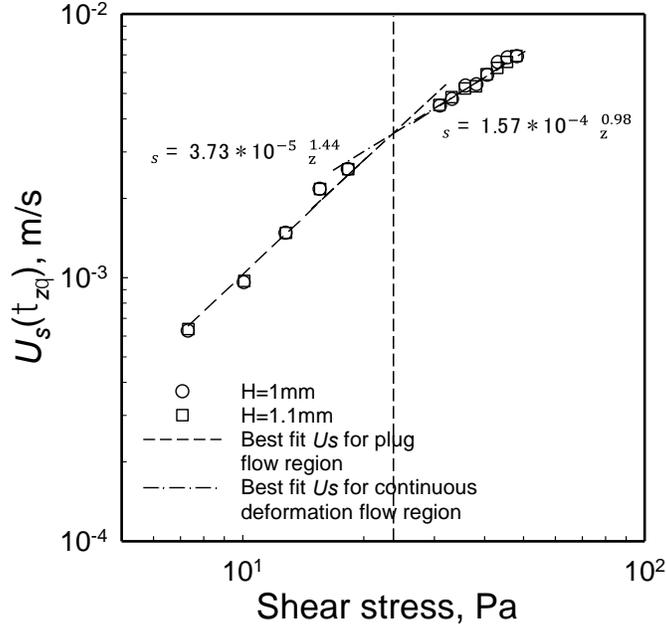

Fig. 11. Wall slip velocity, $U_s$, versus shear stress at the edge, $|\tau_{z\theta}(R)|$, for two gaps from the data collected via PIV experiments [Medina-Bañuelos et al. 2021]. The slope is equal to 3.73*10⁻⁵ m/(Pa$^{1.44}$ s) for plug flow region and 1.573*10⁻⁴ m/(Pa$^{0.98}$ s) for continuous deformation region.

It is interesting to compare the wall slip velocity values determined via the Mooney method with the wall slip velocity data obtained by Medina-Bañuelos *et al.* 2021 employing PIV analysis. The comparisons are shown in Fig. 12. The $\beta$ value varies between $3.73*10^{-5}$ to $5.21*10^{-5}$ and the exponent, $s_b$, ranges from 1.34 to 1.44. The mean values of $\beta$ and $s_b$ from these data are $4.50*10^{-5}$ and 1.39, respectively. Thus, the parameters of the wall slip velocity versus the shear stress relationship stay consistent for the different methods that are utilized.

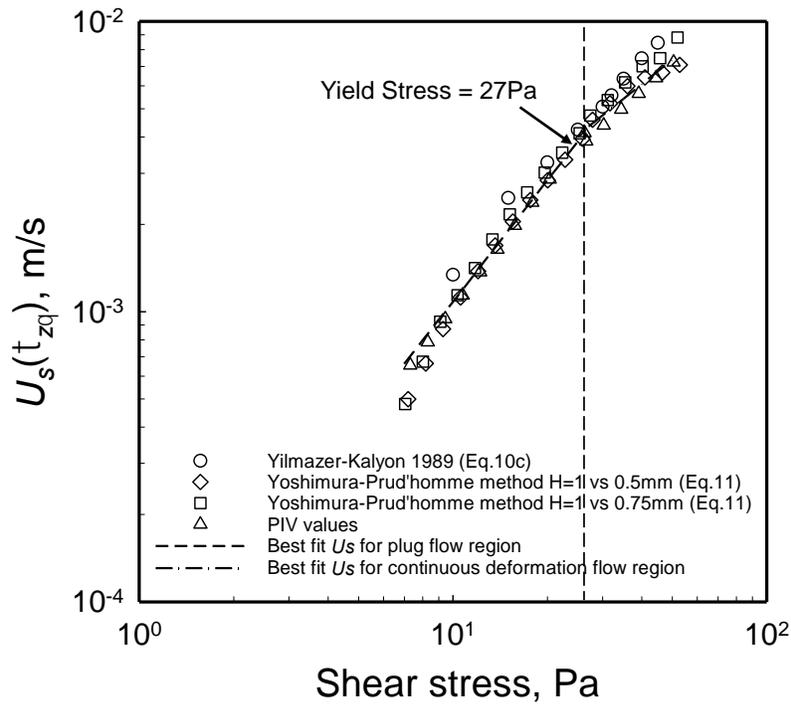

Fig. 12. Slip velocity as a function of shear stress determined using Mooney procedures with best fits for the plug flow and continuous deformation region that were reported in Fig. 11.

To validate the yield stress value obtained with the torque versus the apparent shear rate data one can also probe the relationship between the wall slip velocity and the velocity of the disk driving the steady torsional flow. The ratio of the wall slip velocity over the velocity of the top plate at the edge, i.e., $\dfrac{U_s}{\Omega R}$ versus the shear stress at the edge, $|\tau_{z\theta}(R)|$, is shown in Fig. 13. As indicated earlier, plug flow is indicated when the ratio $\frac{U_s}{\Omega R}$ =0.5 [Kalyon, 2005]. The transition between the plug flow and the continuous deformation flow region is the yield stress of the suspension [Kalyon, 2005]. As shown in Fig. 13 this transition representing the yield stress occurs at around 27 Pa. Thus, the yield stress value determined from wall slip analysis agrees with the yield stress determined from torque versus the apparent shear rate data. As expected, the effect of wall slip is more significant at the smaller gaps, i.e., at greater surface to volume ratios. A third method involving the velocity distributions obtained experimentally, as well as obtained upon computations with the parallel plate analogy, was also applied and again generated a yield stress value close to 27 Pa. This third method will discussed in conjunction with Fig. 15 and Fig. 16.

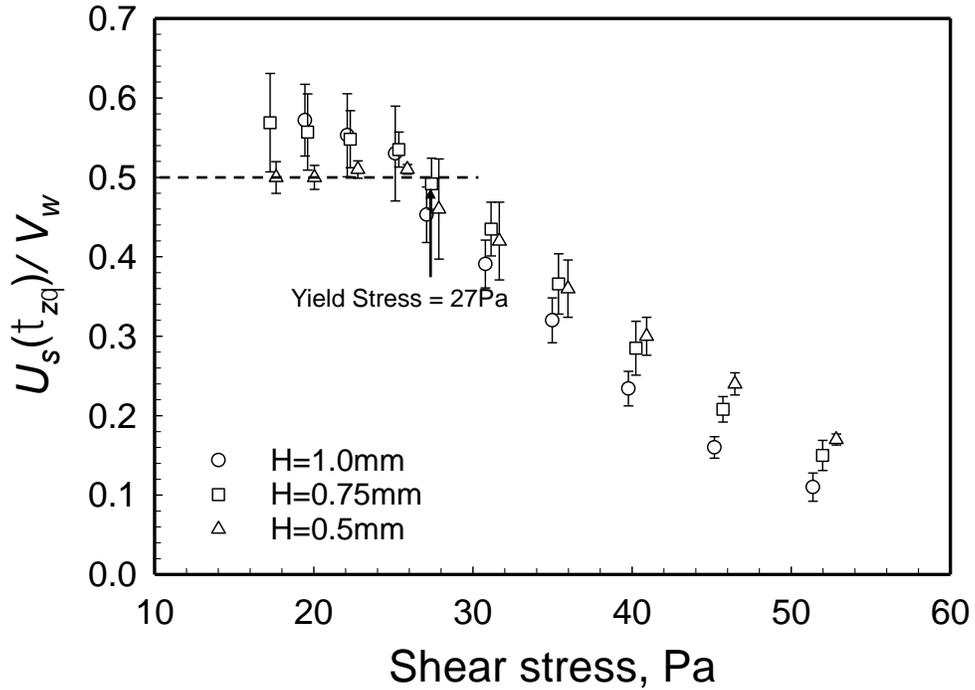

Fig. 13. Wall slip velocity, $U_s$, over the plate velocity, $Vw=\Omega R$, versus shear stress at the edge, $|\tau_{z\theta}(R)|$, from the data collected at three gaps. $\frac{U_s}{\Omega R}$ =0.5 corresponds to plug flow.

The shear stress at the edge versus the true shear rate at the edge for the three gaps are shown in Fig. 14. Since the yield stress value of the hydrogel could be determined from the torque versus the apparent shear rate data directly, the other two parameters of the Herschel-Bulkley equation could be readily obtained from the flow curves, i.e., the shear stress at the edge versus the true (slip corrected) shear rate at the edge. The best fit of the flow curve generated the other two parameters of Herschel-Bulkley Equation as: $m$=3.14 Pa-s$^n$ and $n$=0.54 (Fig. 14). The shear viscosity of the hydrogel used here was characterized earlier using Couette and vane in cup flow and its Herschel-Bulkley parameters were determined in these earlier investigations as $\tau_0$ =27 Pa, $m$=5.5 Pa-s$^n$ and $n$=0.43 [Medina-Bañuelos *et al.*, 2017; Medina-Bañuelos *et al.*, 2019]. The good agreement of the parameters obtained with different viscometric flows is indicative of the robustness of the methodologies used in determining the yield stress and the wall slip velocity versus shear stress behavior of the hydrogel from parallel-disk viscometry.

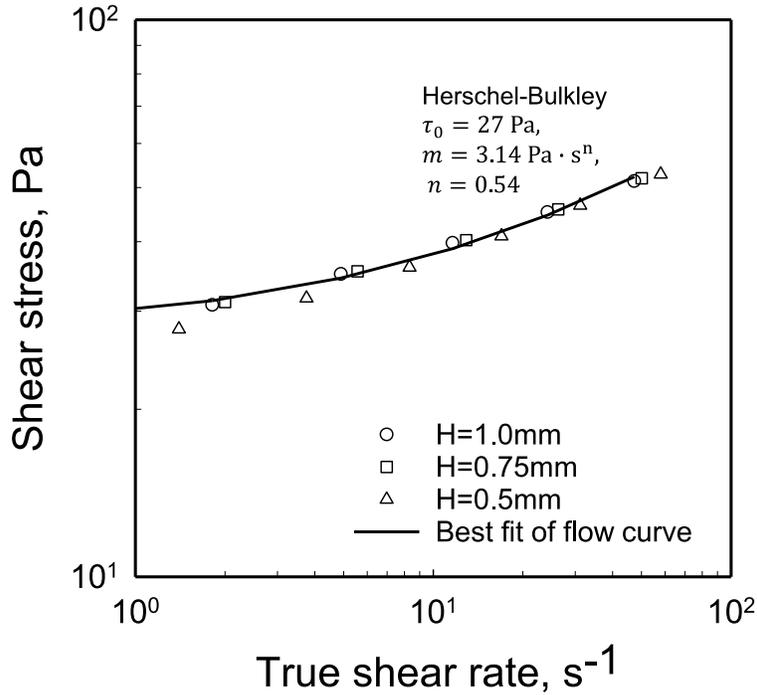

Fig. 14. Shear stress at the edge, $\left|\tau_{z\theta}(R)\right|$, versus the true (slip corrected) shear rate at the edge, $\dfrac{dV_\theta}{dr}(R)=\dot\gamma(R)$, from the data collected at three gaps, *H*. The parameters of the Herschel-Bulkley equation of the hydrogel are yield stress, $\tau_0$ =27 Pa, $m$=3.14 Pa-s$^n$ and $n$=0.54.

Table 1 shows the parameters of the wall slip velocity versus the shear stress for the plug flow and the continuous deformation regions and the parameters of the shear viscosity, employing the viscoplastic Herschel-Bulkley constitutive equation. These parameters were subjected to an additional test involving the predictions of the velocity distributions and torques and their comparisons with the experimental values as discussed next.

Table 1: The parameters of wall slip and shear viscosity.

| Slip velocity versus shear stress | $U_s(R) = \pm \beta \left(-\tau_{z\theta}(R)\right)^{s_b}$ |
|---|---|
| Plug flow region | Continuous deformation region |
| $\beta$= 2.56*10$^{-5}$ m/(Pa$^{s_b}$ s), $s_b$=1.56 | $\beta$= 1.57*10$^{-4}$ m/(Pa$^{s_b}$ s), $s_b$=0.98 |
| Herschel-Bulkley Equation (Equation 2) | |
| $\tau_0$ =27 Pa, $m$=5.5 Pa-s$^n$ and $n$=0.43 | |

The velocity distributions that Pérez-González and co-workers, 2021, collected using the PIV method are shown in Fig. 15(a-f) and Fig. 16(a-h). Fig. 15(a-f) show a set of velocity distributions for different torque, $\Im$, values in the range from 0.2 to 0.64 mN-m. The velocity distributions were obtained for $H$=1.0 mm for each $\Im$ value. The corresponding rotational speed, $\Omega$, values measured were in the range of 0.054 to 0.28 rad/s. Under these conditions the tangential velocity values are constants in between the two plates, i.e., the flow of the hydrogel is plug flow. This is consistent with how viscoplastic fluids flow when the imposed shear stress is less than the yield stress of the fluid (Fig. 15(a-f)). As an example, the measured values of the slip velocities for a $\Im$ value of 0.2 mN-m were $U_s(R_m,0) = 6.28*10^{-4}$ m/s and $U_s(R_m,H) = -6.22*10^{-4}$ m/s at $H$=1.0mm, to indeed generate the following as required for plug flow:

$$U_s(R_m,0) = -U_s(R_m,H) = \Omega R_m / 2 \qquad (13)$$

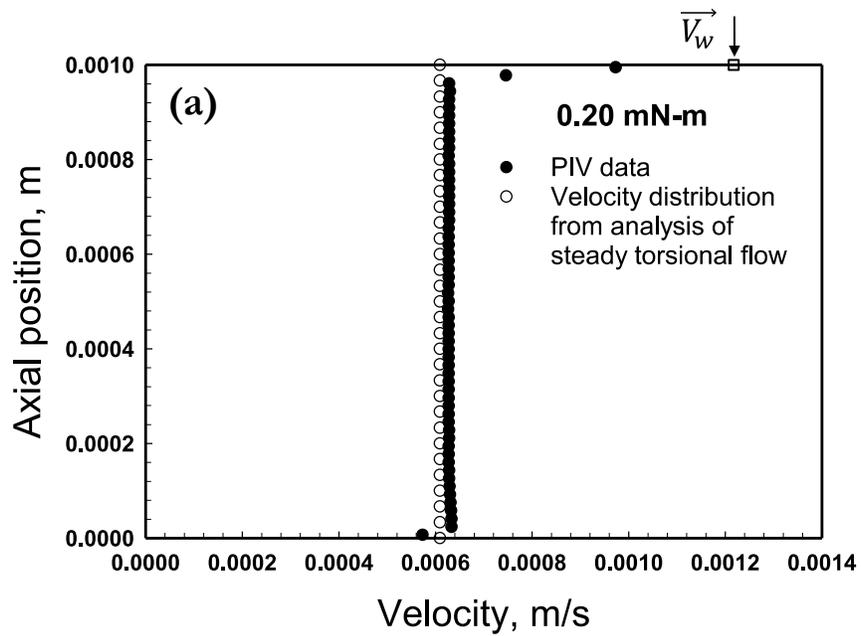
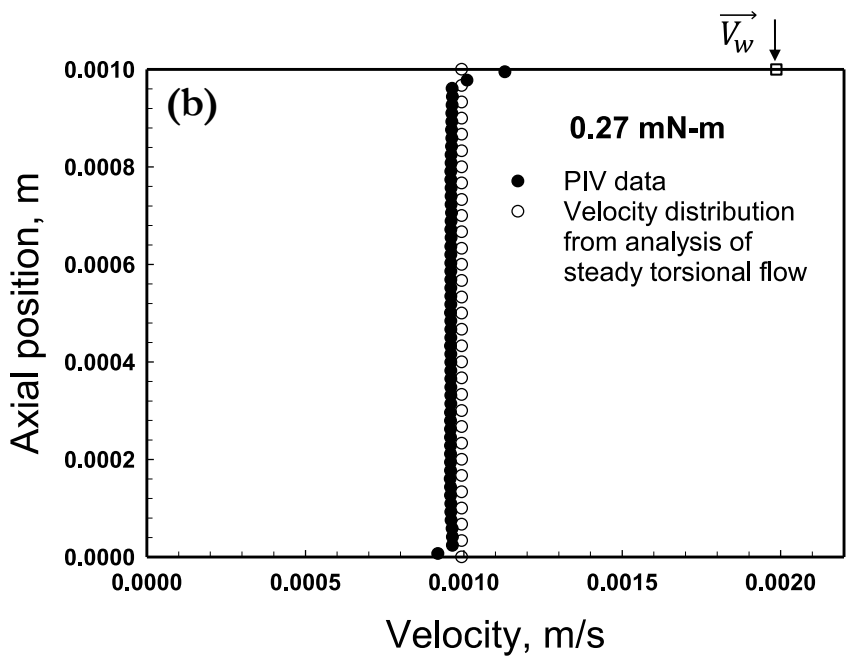

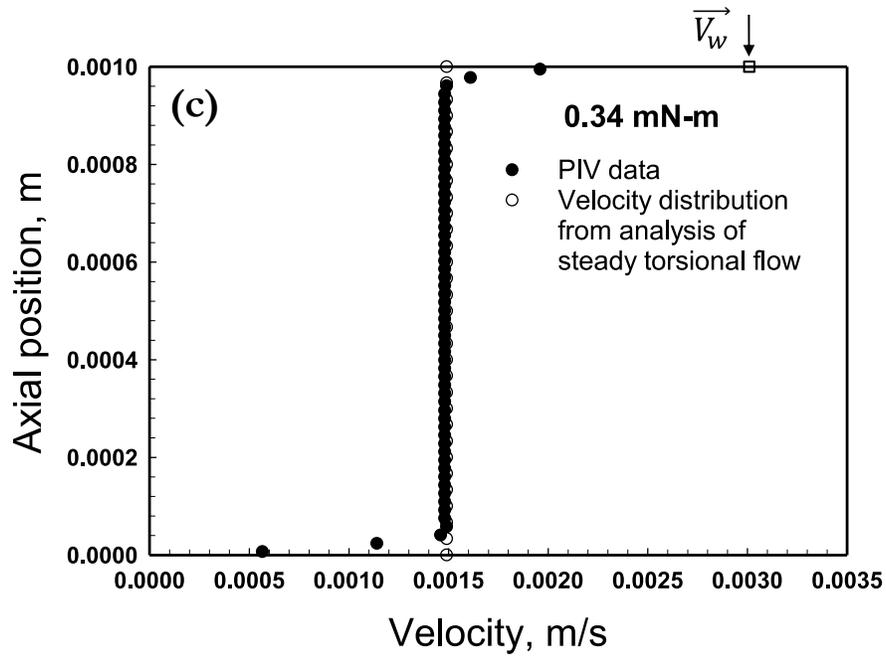

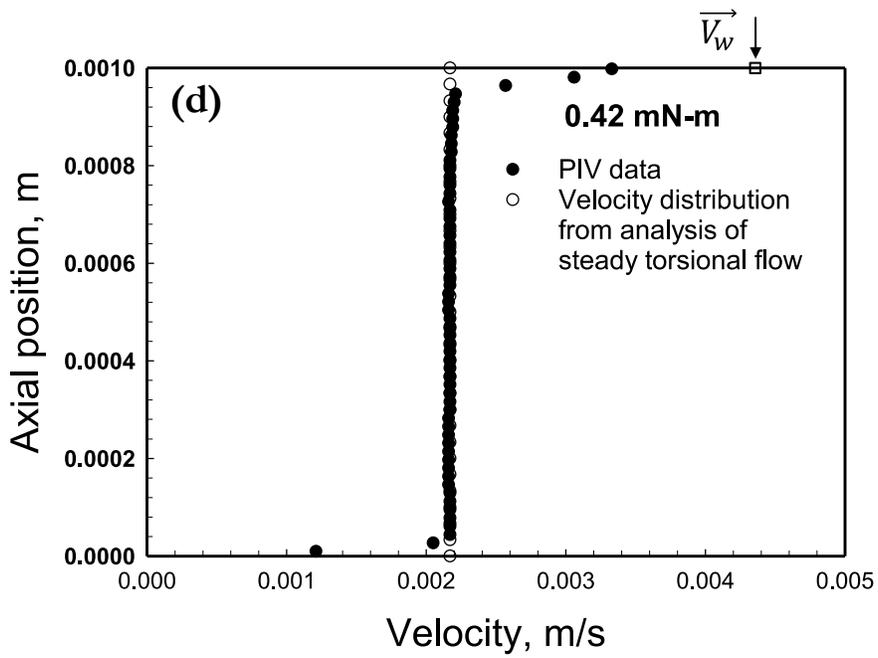

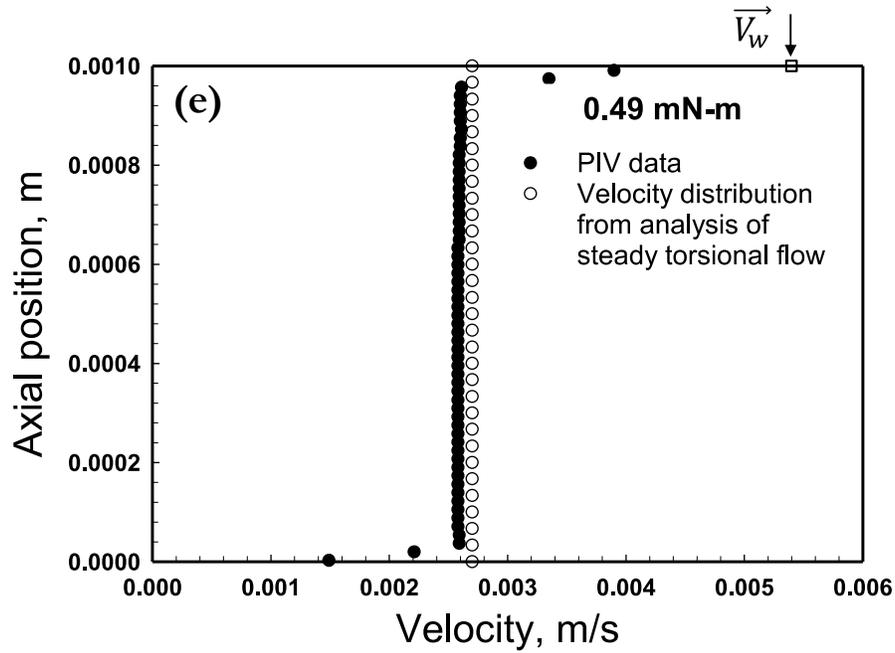

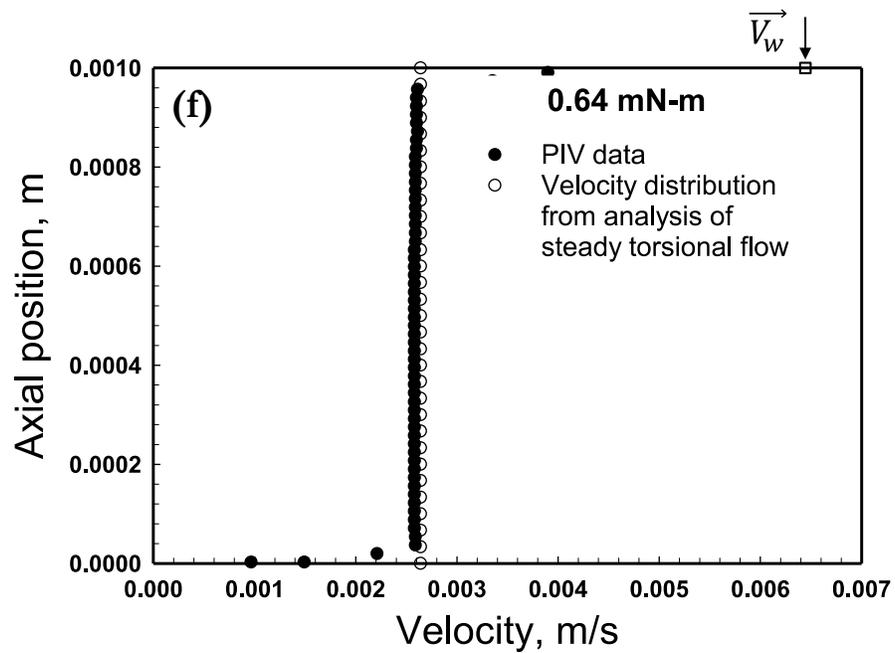

Fig. 15. Velocity distributions in steady torsional flow using parallel disks for various $\Im$ values in the range from 0.20 to 0.64 mN-m for the gap, $H$=1.0 mm. The PIV data are from Medina-Bañuelos *et al.*, 2021.

On the other hand, the calculations and the experimental data for the velocity distributions corresponding to the torque range of 0.93 to 1.52 mN-m indicate that when the shear stress at a given radial position exceeds the yield stress there is continuous steady deformation (Fig. 16(a-h)). It is observed that for $\left|\tau_{z\theta}(r)\right| > \tau_0$ the tangential velocity, $V_\theta(r)$ increases linearly with axial distance, $z$ and the deformation rate, $\frac{dV_\theta}{dz}(r)$ increases with increasing rotational speed, $\Omega$. Similar to the plug flow case apparent wall slip plays a key role.

Fig. 16(a-h) show various velocity distributions for different torque, $\Im$ values in the range from 0.93 to 1.45 mN-m for $H = 1.0$ mm predicted with the parameters of the Herschel-Bulkley equation and wall slip versus shear stress relationship. The parallel plate analysis (Eq. 7) was applied and the torque, $\Im$ was determined via numerical integration (Eq. 9a). The corresponding rotational speed, $\Omega$ values in this case varied from 0.45 to 3.0 rad/s. It can be seen that the shape of the velocity distributions changed, as compared to those in Fig. 16(a-h), from a constant value throughout the gap, plug flow, to a linearly increasing function of $z$, indicating that the yield stress has been surpassed and the hydrogel is undergoing continuous shear deformation under the condition of the torque surpassing the critical torque, i.e., $\Im \geq \Im_c$, and hence $\left|\tau_{z\theta}(r)\right| > \tau_0$. As expected the distributions in Fig. 16(a-h) do not provide evidence of yielding within the gap that delineates the regions of plug and shear flow within the gap, as observed in the cases of Couette (or vane in cup), capillary or axial annular flows of viscoplastic fluids [Aktas *et al.*, 2014; Ortega-Avila *et al.*, 2016; Medina-Bañuelos *et al.*, 2017, Medina-Bañuelos *et al.*, 2019]. The yielding within the gap may be seen in those flows when $\left|\tau_{z\theta}(r)\right| > \tau_0$ at some position within the gap because the shear stress distribution is non-homogeneous across the gap. However, during steady torsional flow the shear stress $\tau_{z\theta}(r)$ is constant at any gap, regardless of the axial position. The velocity distributions predicted via our MATLAB code based on Herschel-Bulkley parameters and wall slip parameters (Equations (7a and 7b)) are shown together with the experimental velocity distributions. The agreement between the experimental distributions and the numerical simulation results suggest that the parallel plate analysis is satisfactory to represent the flow and deformation occurring in steady torsional flow and that the parameters of shear viscosity and wall slip are accurate.

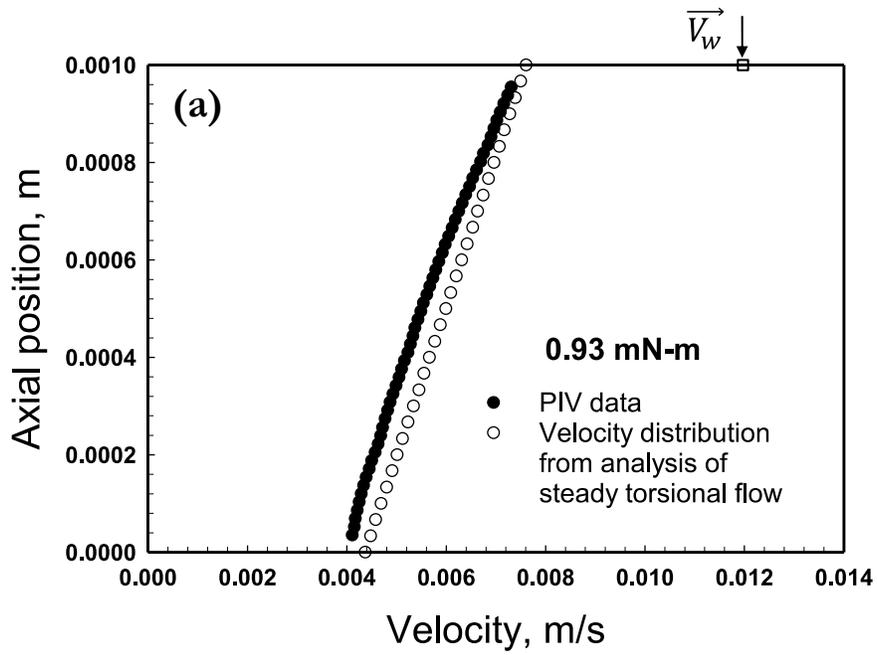

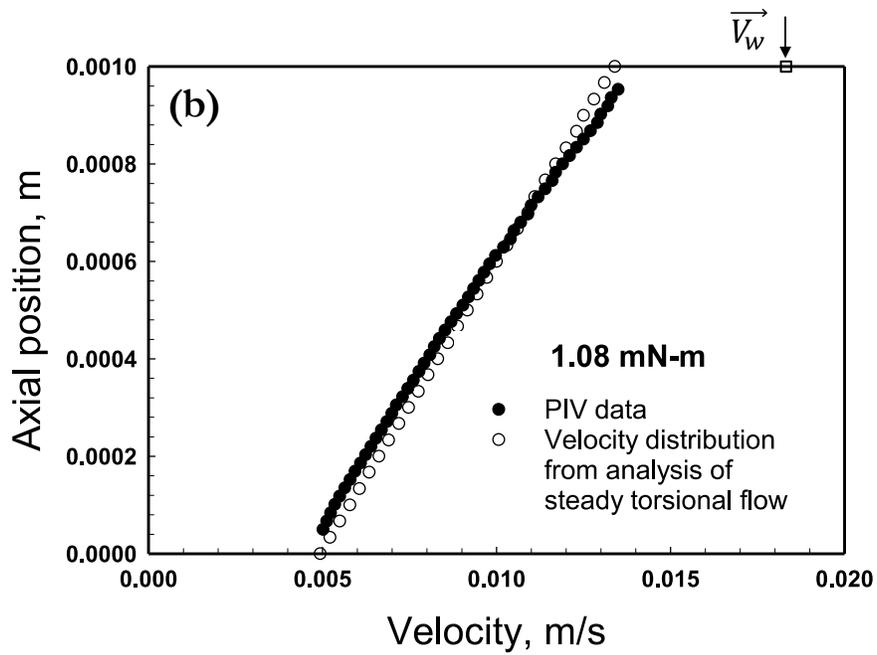

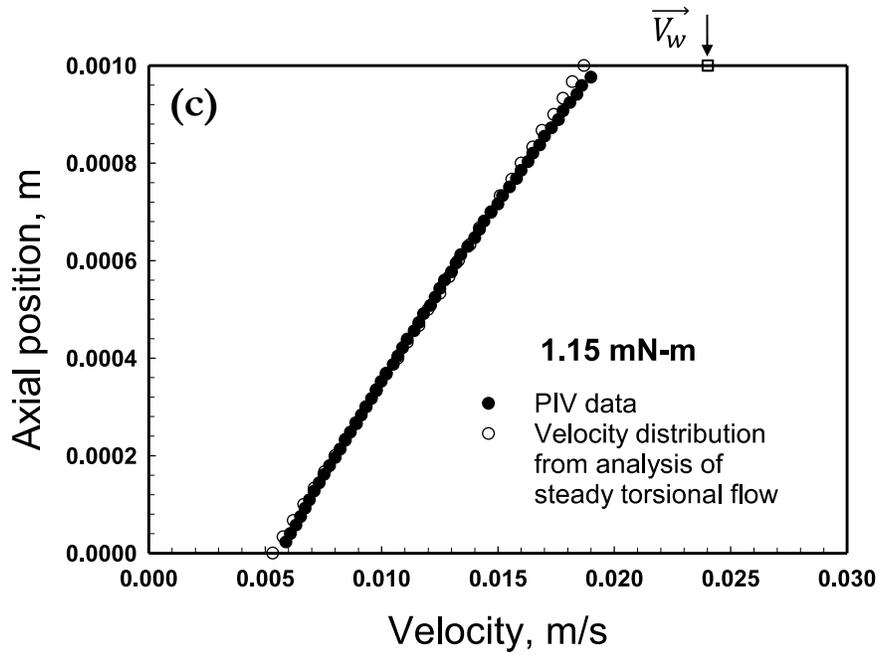
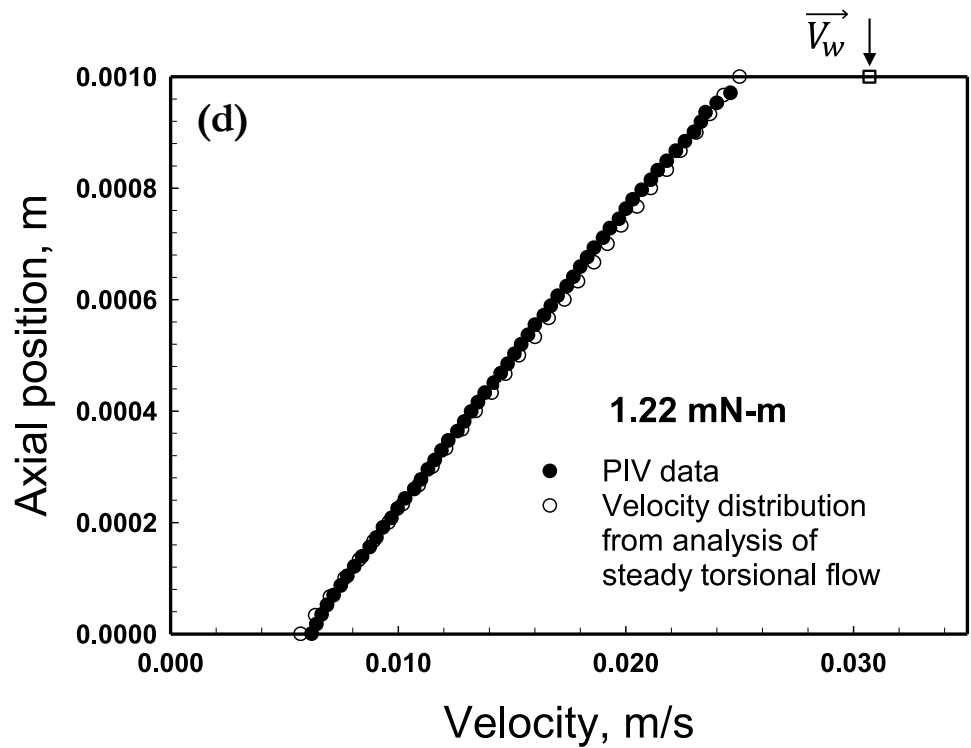

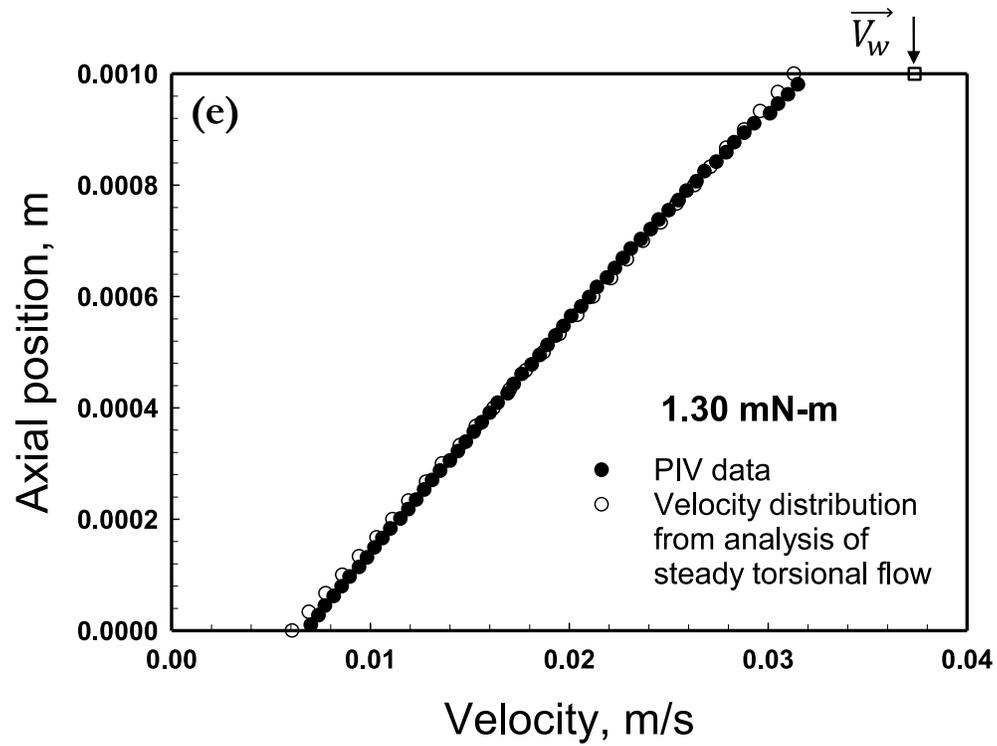

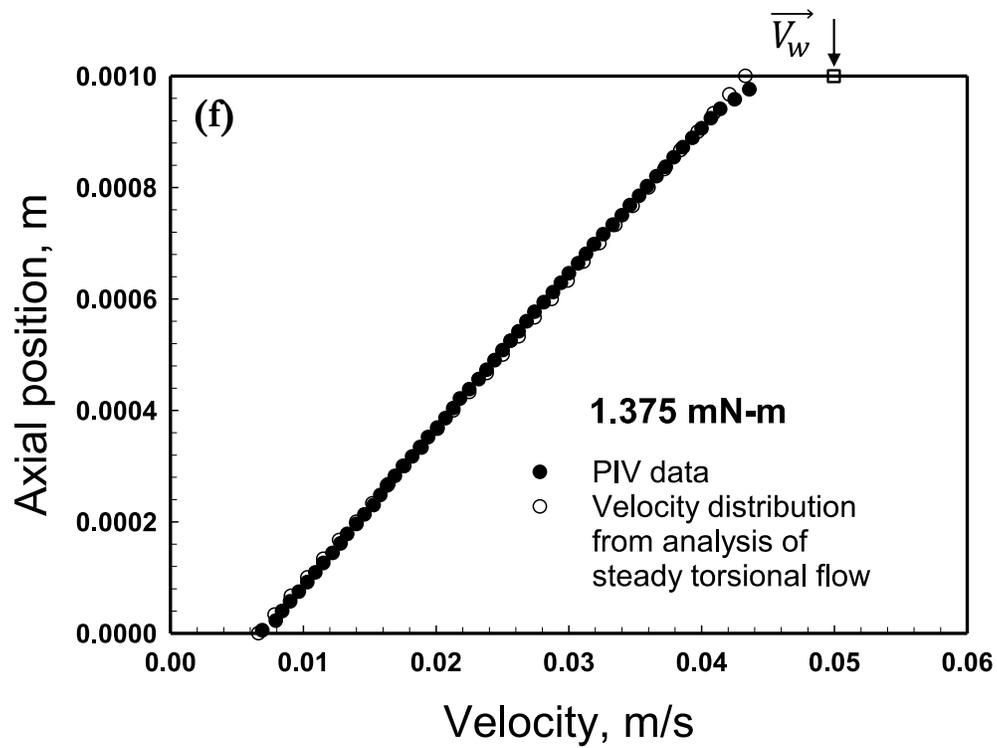

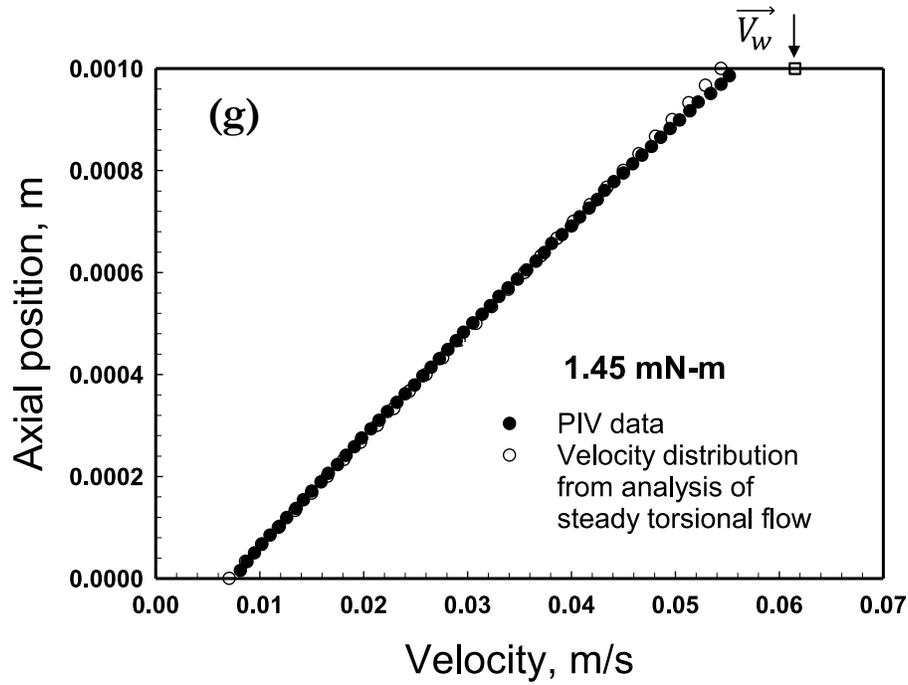

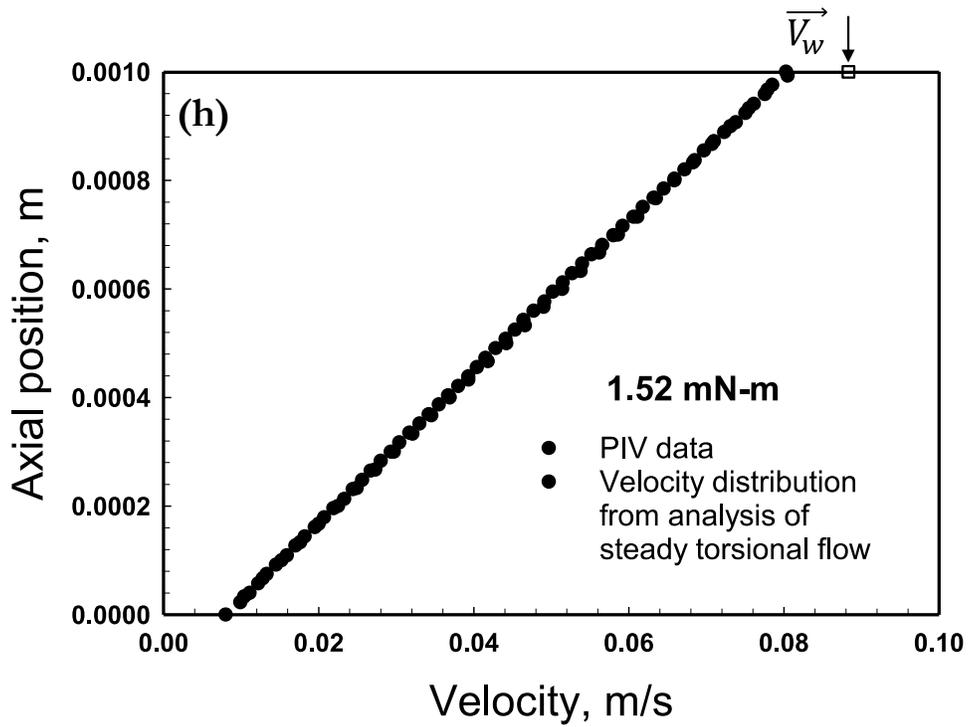

Fig. 16. Velocity distributions in steady torsional flow using parallel disks for various torque, $\Im$, values in the range from 0.93 to 1.52 mN-m for the gap, $H$=1.0 mm. The PIV data are from Medina-Bañuelos *et al.*, 2021.

Can the velocity distributions (Fig. 15 and 16) be used to bracket the yield stress value? The plug flow observed in Fig. 15f occurs at a shear stress of 24 Pa whereas the continuous deformation profile shown in Fig. 16a occurs at a shear stress of 30 Pa. Therefore the yield stress of the hydrogel is in between 24 to 30 Pa, consistent with the yield stress obtained using the torque versus apparent shear rate data (Fig. 6 and 7) which had identified the same shear stress range for the yield stress.

To test further the accuracy of the parameters of wall slip velocity versus the shear stress relationship and the shear viscosity material function of the hydrogel, the torques under different conditions were also solved by numerical integration. For this the parallel plate approximation was again used, i.e., Equation (7), which was solved incrementally in the radial direction for each set of disk rotational speed, $\Omega$, and gap, $H$, via numerical integration using the MATLAB code. This additional step of the prediction of the torques and their comparisons with the experimental torque values of Medina-Bañuelos *et al.*, 2021, would provide an additional assessment of the accuracy of the parameters of shear viscosity and wall slip characterized using the torque versus apparent shear rate data. The first step in this procedure is the determination of the shear stress distributions as a function of the radial position, r.

The shear stress distributions, $\tau_{z\theta}(r)$, for various apparent shear rates as a function of the radial position, $r$, (at gap, $H$=1mm), are shown in Fig. 17. The shear stress increases monotonically with increasing $r$, reaching a maximum at the edge of the disk. There is a significant difference in the shear stress distributions obtained under the plug flow conditions (apparent shear rates in the 1.3 to 7.1 s$^{-1}$) and the continuous deformation conditions (apparent shear rates in the 11.4 to 97.5 s$^{-1}$). The yield stress range of 24-30 Pa clearly delineates the shear stress distributions into the expected two zones related to plug flow and the continuous deformation conditions.

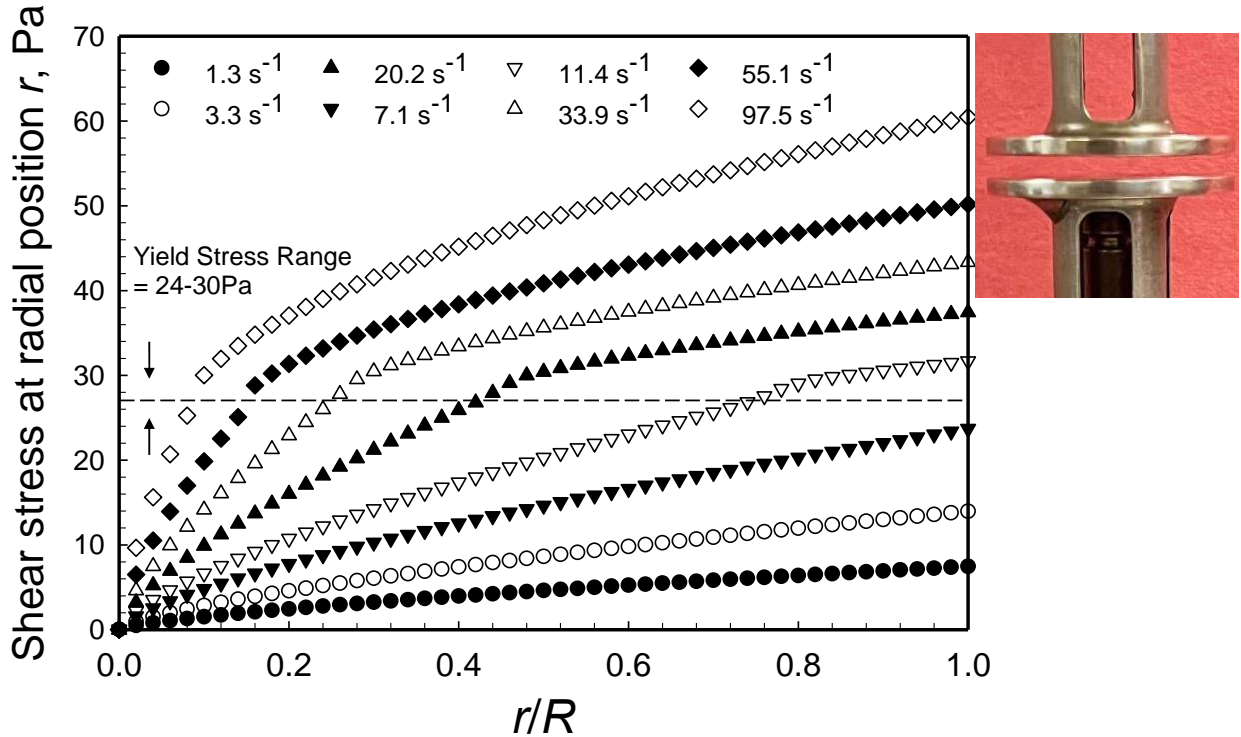

Fig. 17. Radial distributions of apparent shear stress, $\left|\tau_{z\theta}(r)\right|$ in steady torsional flow between two disks for the range of apparent shear rates at edge, $\dot{\gamma}_{aR} = \Omega R / H$, between 1.3 to 97.5 s$^{-1}$ ($H$=1 mm).

Upon the calculation of the shear stress distribution, $\tau_{z\theta}(r)$, the torque, $\Im$, at each apparent shear rate was obtained via numerical integration, i.e., $\Im(H,\Omega) = 2\pi \int_0^R \left(-\tau_{z\theta}(r)\right) r^2 dr$ such as by using $\Im(r) = 2\pi \left(-\tau_{z\theta}(r)\right) r^2 \Delta r$. The typical $\Delta r$ values were around 0.0005m for $R$=0.025 m, i.e., $\Delta r/R$ was 0.02. The effect of the choice of $\Delta r$ was probed by systematically changing $\Delta r$ in the 0.0005 to 0.00005 m range. As shown in Appendix A (Fig. A-1) the torque results converge and $\Delta r$ is no longer a factor when $\Delta r$ is smaller than 0.0005 m.

The comparisons of the converged torques obtained via numerical integration employing the characterized parameters of wall slip and shear viscosity (Table 1), with the experimental torque values are shown in Fig. 18. There is excellent agreement between the experimental torque values and those that were numerically determined (Fig. 18). The excellent agreement is an additional testament that the parallel plate approximation made under the condition of $H<<R$ (lubrication assumption) is acceptable and that the parameters of the shear viscosity and the slip velocity behavior are accurate. In the following section a second geometry, i.e., the cone-and-plate flow, will also be analyzed and its utility versus the parallel disk viscometry will be elucidated.

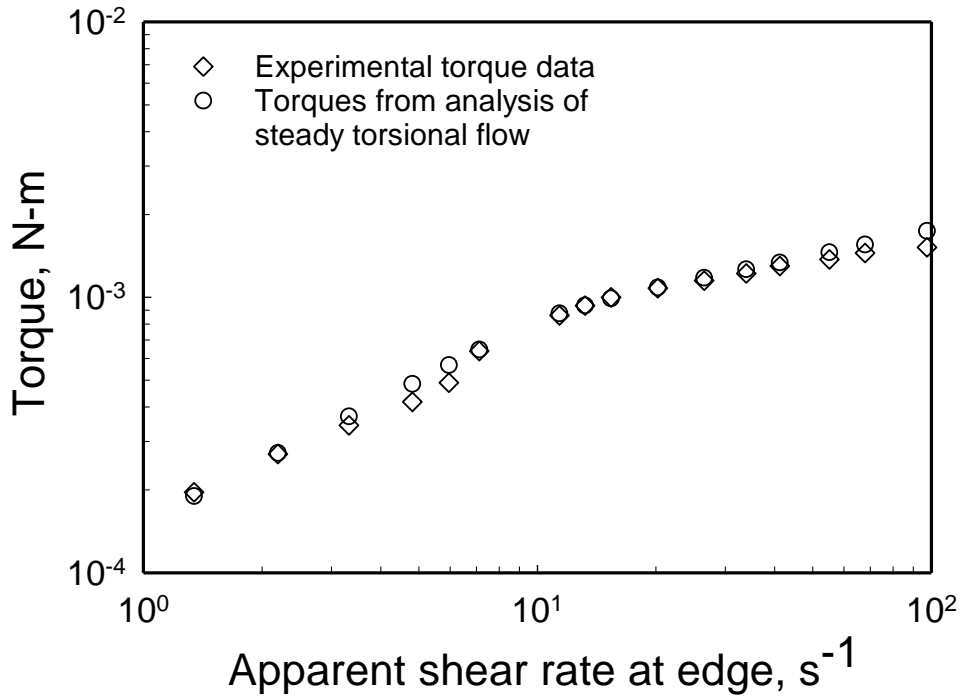

Fig. 18. The comparisons of the converged torques obtained via numerical integration employing the characterized parameters of wall slip and shear viscosity with the experimental torque values.

**Cone-and-plate flow:**

The cone-and-plate configuration is shown in Fig. 4b and Fig. 19a. The sample is sandwiched in between a rotating cone and a stationary disk (plate). Either the torques are systematically imposed and the corresponding steady rotational speeds are determined or the rotational speeds are systematically varied and the corresponding steady torques are determined.

The cone-and-plate flow is considered to be a useful viscometric flow configuration since for simple fluids the shear rate and shear stress are independent of the radial position. This can be seen from the following. It is considered that the cone angle $\alpha$ is so small that a lubrication approximation may be applied to the flow in the gap [Bird *et al.*, 1987]. As also discussed for the steady torsional flow between two parallel disks, the lubrication approximation considers the flow to be locally occuring in between two parallel disks (plates). For the cone and plate system, at a distance $r$ from the cone apex, the velocity of the cone is $\Omega r$, and the plate-cone separation is $H(r) = r\sin\alpha = r\alpha$. Accordingly, the velocity profile can be approximated as the following employing the spherical coordinate system ($r$, $\theta$, $\phi$) as depicted in Fig. 19a [Bird et al., 1987]:

$$v_\phi = \Omega r \left( \frac{(\pi/2) - \theta}{(\pi/2) - \theta_1} \right) \tag{14}$$

The $\theta\phi$-component of the rate of deformation tensor is then:

$$\dot{\gamma}_{\theta\phi} = \frac{\sin\theta}{r} \frac{\partial}{\partial \theta}\left(\frac{v_\phi}{\sin\theta}\right) \doteq \frac{1}{r}\frac{\partial}{\partial \theta} v_\phi = -\frac{\Omega}{\alpha} \tag{15}$$

The approximation made here is that $\theta$ is so close to $\pi/2$ that $\sin\theta$ can be taken to be unity; this should be an excellent approximation. We see from Equation 15 that the shear rate $\dot{\gamma}_{\theta\phi}$ is constant throughout the cone-plate gap. As noted by Bird and co-workers [Bird et al., 1987] imposing a shear rate that is independent of the rheological behavior of the fluid is an excellent way for the characterization of the shear rate-dependent viscosity behavior. The torque can be determined as $\Im = \int_0^{2\pi} \int_0^R \tau_{\theta\phi}\big|_{\theta=\pi/2} r^2 dr d\phi$ [Bird et al., 1987]. Since $\dot{\gamma}_{\theta\phi}$ is constant for any $r$, the shear stress, $\tau_{\theta\phi}$ will also be a constant, so that $\tau_{\theta\phi} = \frac{3\Im}{2\pi R^3}$ and hence the shear viscosity becomes $\eta = \frac{\tau_{\theta\phi}}{-\dot{\gamma}_{\theta\phi}} = \frac{3\Im\alpha}{2\pi R^3 \Omega}$ [Bird et al., 1987]. For shear rate dependent generalized Newtonian fluids, for example, the power-law expression, the shear viscosity material function, $\eta(\dot{\gamma}_{\theta\phi}) = \frac{\tau_{\theta\phi}(\dot{\gamma}_{\theta\phi})}{-\dot{\gamma}_{\theta\phi}}$ and $\tau_{\theta\phi}(\dot{\gamma}_{\theta\phi}) = -m \cdot \dot{\gamma}_{\theta\phi}^n$.

Therefore, for generalized Newtonian fluids the shear viscosity of the fluid in the gap can be determined on the basis of the geometry, i.e., the radius of the cone and the disk, $R$, and the cone angle $\alpha$, and the measured values of torque $\Im$ at systematically varied angular velocity values, $\Omega$. Thus, to summarize for generalized Newtonian fluids that are not viscoplastic and which do not exhibit wall slip the cone-and-plate flow provides the benefits of having an *a priori* known (Eq. 15) shear rate which remains constant within the gap in between the cone and the plate (regardless of the radial position) and enables the facile characterization of the shear viscosity material function.

Would the shear rate and the shear stress remain constant over the radial distance, $r=0$ to $r=R$ for viscoplastic fluids with wall slip? This is tested for the hydrogel of this investigation, based on the lubrication assumption and resulting parallel plate approximation, i.e., via Equations (7a) and (7b). The numerically computed shear stress, wall slip velocity and shear rate distributions are shown in Fig. 19-21 for a gap of 0.00255 m, radius, $R$, of 0.025 m, and for various rotational speeds of the cone, $\Omega$. Fig. 19b indicates that the shear stress is not a constant within the gap and monotonically increases as r increases, and similar to the distributions of the parallel plate configuration the yield stress delineates the shear stress behavior of the hydrogel below and above its yield stress value, i.e., 27±3 Pa.

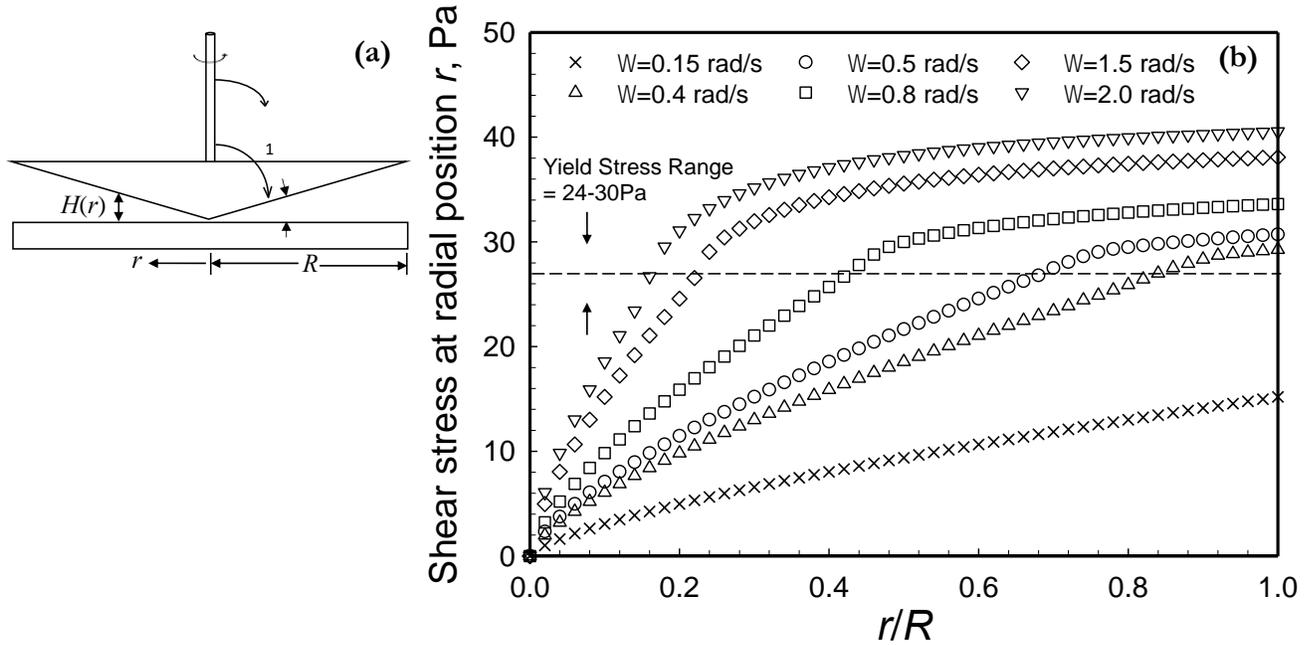

Fig. 19. The cone-and-plate viscometer (a) and the radial distributions of shear stress, $\left|\tau_{z\theta}(r)\right|$ obtained for cone-and-plate flow for rotational speeds, $\Omega$, that are in the range of of 0.15 to 2.0 rad/s (the cone angle, $\alpha$, is 0.1 rad).

The wall slip velocity over the velocity of the moving wall versus the radial position, $r$, in cone-and-plate flow is shown in Fig. 20. The ratio of the slip velocity over the wall velocity reflect the occurrence of plug flow when the shear stress is less than the yield stress, i.e., $U_s/V_w = 0.5$ [Kalyon, 2005]. The ratio decreases with increasing $r/R$ as the hydrogel is deformed continuously $\left|\tau_{z\theta}(r)\right| > \tau_0$. For different rotational speeds, $\Omega$, the radial location at which the shear stress reaches the yield stress of the hydrogel are also indicated in Fig. 20.

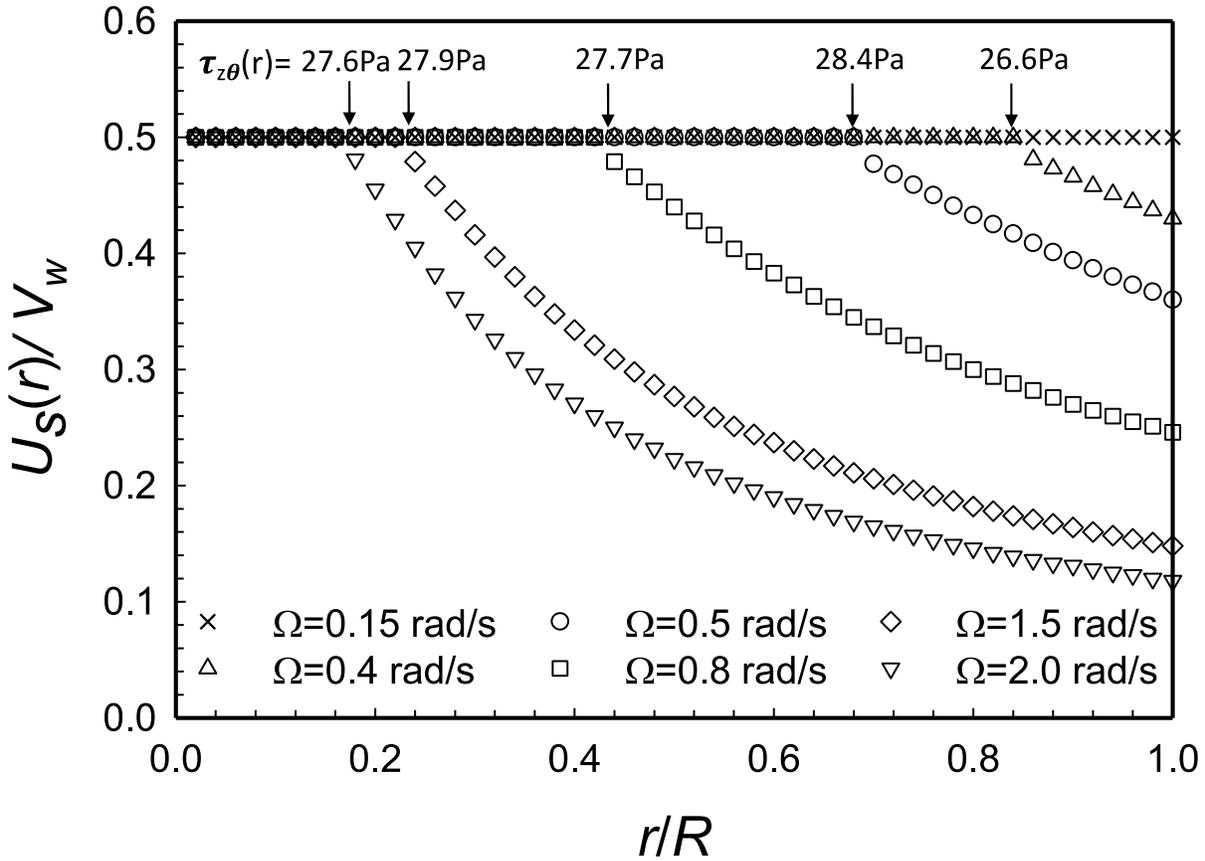

Fig. 20. Radial distributions of wall slip velocity, $U_s$, over the plate velocity ($\Omega R$) in cone and plate flow for the rotational speeds, $\Omega$, of 0.15 to 2.0 rad/s (the cone angle, $\alpha$, is 0.1 rad).

The true shear rates obtained from Equation (10) as a function of $r/R$ are shown in Fig. 21. The transition between the plug flow and the continuous deformation occurs when the $\left|\tau_{z\theta}(r)\right| > \tau_0$. The progression of the location at which the shear stress reaches the yield stress of the hydrogel are also indicated. Overall, there is no deformation but only solid plug flow motion for shear stresses that are less than the yield stress, i.e., $\left|\tau_{z\theta}(r)\right| \leq \tau_0$ and continuous deformation is indicated for shear stresses that exceed the yield stress, i.e., $\left|\tau_{z\theta}(r)\right| > \tau_0$. It can be assumed that the transition in the radial direction in between the plug flow motion and continuous deformation zone occurs over a range of shear stresses (a relatively small range of shear stresses) which are in the neighborhood of the yield stress value so that the fluid does not fracture into two parts in the radial direction.

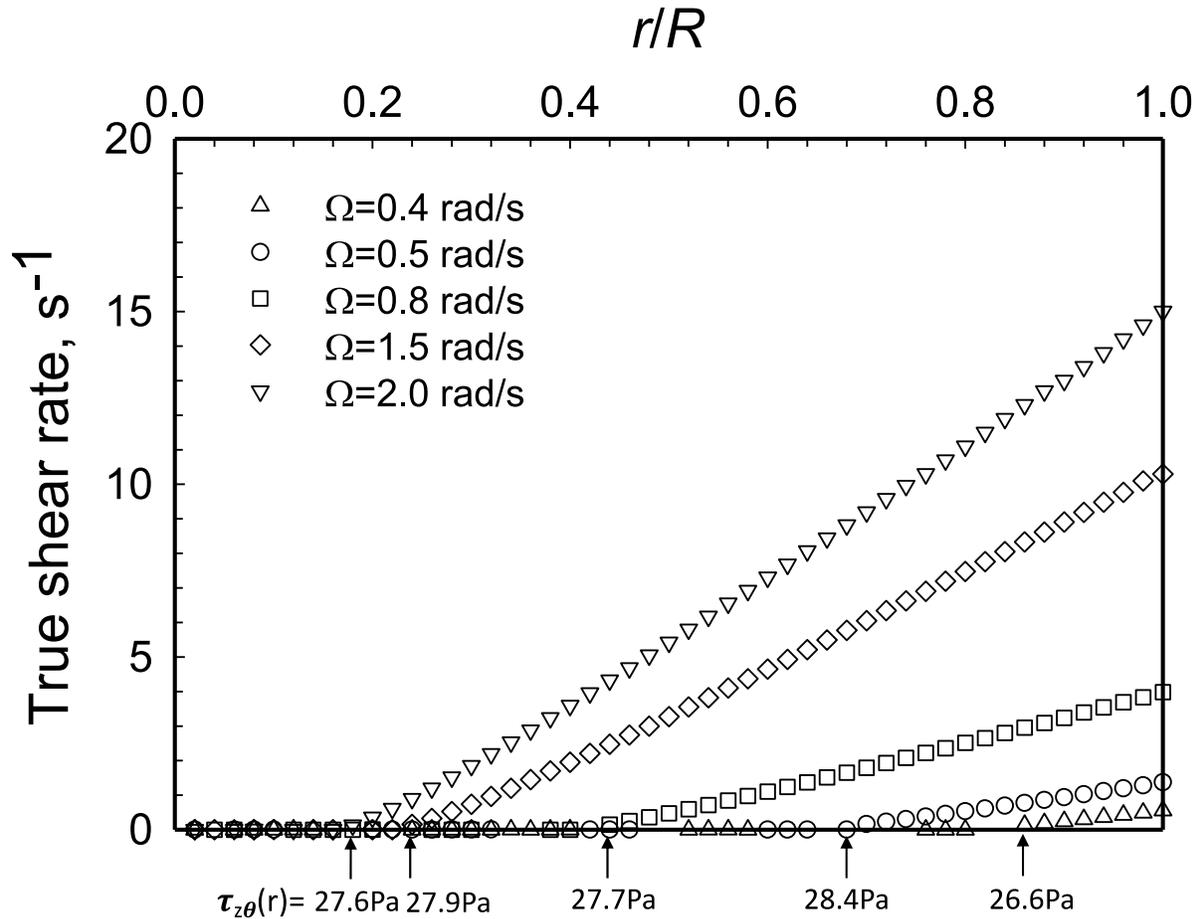

Fig. 21. Radial distributions of true shear rate, $\frac{dV_\theta}{dr}(r) = \dot{\gamma}(r)$, in cone-and-plate flow of the hydrogel for the rotational speeds, $\Omega$, in the range of 0.15 to 2.0 rad/s (the cone angle, $\alpha$, is 0.1 rad).

The results shown in Fig. 19 and Fig. 21 suggest that both the shear rate and the shear stress change as a function of $r$ for the hydrogel. Obviously both viscoplasticity and the apparent wall slip play significant roles. Since the flow in between the cone and the plate is non-homogeneous flow and neither the shear stress nor the shear rate can be known *a priori*, there is no particular advantage to using the cone-and-plate flow for the characterization of the shear viscosity material function of viscoplastic fluids. Furthermore, the additional weakness of cone-and-plate flow, i.e., the presence of a relatively small gap at the apex of the cone that can capture particle clusters and thus erroneously give rise to an artificial normal force (due to particle clusters being compressed in the axial direction) should be considered. Finally, there is also the difficulty of the determination of the yield stress and other parameters of the shear viscosity from cone-and-plate flow for viscoplastic fluids with wall slip; that is there is no straightforward method to convert torques to shear stresses. Thus, overall the parallel-disk viscometry should be preferable over the cone-and-plate flow configuration for the characterization of the shear viscosity and wall slip behavior of viscoplastic hydrogels.

**Conclusions:**

The flow and deformation behavior of hydrogels is central to many of the applications that they are subjected to in myriad areas that are as diverse as biomedical devices, hydraulic fracturing, foodstuffs and personal care products. It is very important to be able to characterize reproducibly and accurately the rheological behavior of hydrogels using steady simple shear flows so that the flow and deformation behavior of the hydrogel can be readily understood and, if necessary, further tailored to the requirements of the application at hand. It is the viscoplasticity and the slip at the wall behavior of the hydrogels that render such characterization and tailoring difficult. Here, one of the simplest rheological characterization methods, the parallel-disk viscometry, i.e., the steady torsional flow using parallel disks, is analyzed in detail and compared with the cone-and-plate flow. The analysis was carried out on a Carbopol® hydrogel (0.12% by weight poly(acrylic acid)). It was shown that the parallel-disk viscometry demonstrates that the hydrogel is viscoplastic with a yield stress in the range of 24-30 Pa (mean 27 Pa). This yield stress value is consistent with earlier investigations that relied on other types of viscometric flows including Couette flow in between two concentric cylinders and vane in cup flow. It is demonstrated that the method for the determination of the yield stress value of viscoplastic fluids in general and the hydrogel in particular, via parallel-disk viscometry is very simple to implement and only relies on the collection of the torque versus the rotational speed data. It is shown that once the yield stress value is determined that other parameters of viscoplastic constitutive equations, including the Herschel-Bulkley fluid, can be determined following the analysis of the wall slip behavior of the hydrogel.

The analysis of the wall slip velocity versus the shear stress behavior of the hydrogel was carried out in conjunction with the apparent slip mechanism, i.e., the formation of a particle free binder-rich zone at the two walls of the viscometer. Such an apparent slip mechanism is widely encountered for concentrated suspensions and gels. It is determined that the mechanisms for the formation of the apparent slip layer thickness are different when the hydrogel is undergoing plug flow ($|\tau_{z\theta}(R)| \leq \tau_0$) or continuous deformation flow, i.e., $|\tau_{z\theta}(r)| > \tau_0$. The results indicate that the apparent slip layer consists of only water for the continuous deformation flow region. However, a complex behavior is observed for the plug flow region. PAA chains are attached firmly to the particles at one end and are free, "dangling", to rotate and orient at their free end. It is hypothesized that the dangling PAA chains could be present within the apparent slip layer to render the apparent slip layer non-Newtonian. At high shear stress and shear rates of the continuous deformation region, the dangling chains would orient along the streamlines, and clear away from the apparent slip layer, thus leaving only water to constitute the slip layer.

Following the determination of the yield stress of the hydrogel from the torque versus the apparent shear rate data, the application of systematic changes in the surface to volume ratio of the parallel disk rheometer allows the determination of the wall slip velocity versus the shear stress relationship, followed by the consistency index, *m*, and shear sensitivity exponent, *n*, of the Herschel-Bulkley fluid. The parameters of the shear viscosity and the apparent wall slip, thus obtained, were tested by being used for the predictions of the velocity distributions and the torques obtained under different flow conditions (employing a simple parallel plate approximation in conjunction with the lubrication assumption) and were compared with the data of Pérez-González and co-workers [Medina-Bañuelos *et al.*, 2021]. The excellent agreements between the predicted and experimentally determined torque values and the velocity distributions are testaments to the reliability of the parameters and the suitability of the parallel plate flow approximation-based methods for the analysis of the parallel disk viscometry data.

The numerical analysis employing the parameters of the shear viscosity and wall slip velocity versus the shear stress relationship were also applied to the cone-and-plate flow of the hydrogel. Although the torques predicted again agreed very well with the experimental torque values obtained for cone-and-plate flow it was not possible to determine the parameters of shear viscosity and wall slip in a straightforward fashion, as could be done for parallel disk viscometry. It was shown that during the flow in between the cone and the plate neither the shear stress nor the shear rate remain constant. The shear stress and shear rate are functions of the radial distance, r, affected by the viscoplasticity and wall slip behavior, with a plug flow zone found surrounding the axis of symmetry. Thus, the cone and plate geometry does not offer any advantages over the parallel disk viscometry, and is not recommended to be applied for the rheological characterization of viscoplastic fluids, including viscoplastic hydrogels.

**Acknowledgments:**

We acknowledge with gratitude the permission given by Prof. Pérez-González and co-workers for the use of their torque versus apparent shear rate data and the velocity distributions from PIV experiments from parallel-disk viscometry [Medina-Bañuelos *et al.*, 2021]. We also thank Elsevier for their permission to include Figures 2 and 3 in our manuscript.

**Appendix A**

Fig. A-1 shows that the torque values, determined numerically using the parameters of wall slip velocity and the shear viscosity of the hydrogel (Table 1) in conjunction with the parallel plate approximation, is independent of the $\Delta r$, if it is small enough. Torque values were obtained via MATLAB under a gap of 1 mm and rotational speed of 2 rad/s.

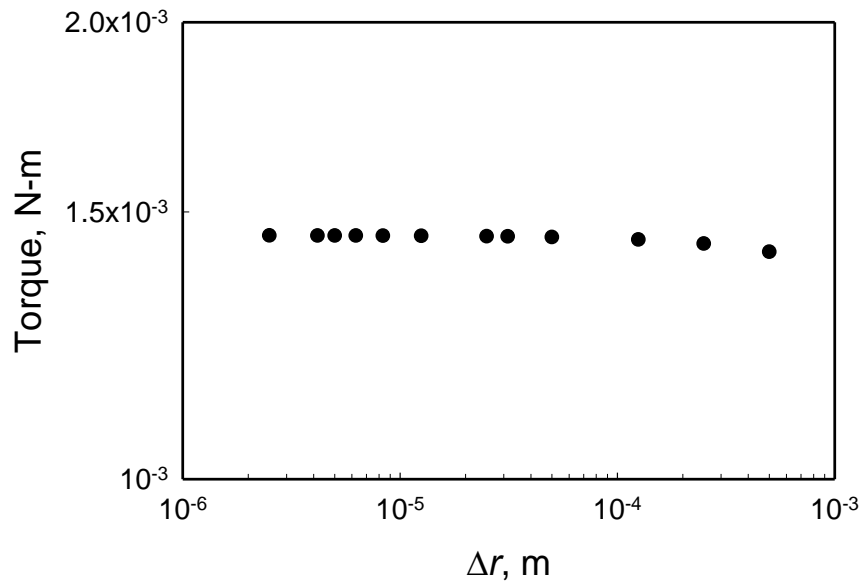

Fig. A-1. The effects of the radial distance increment used on the torques obtained upon numerical integration for the conditions of $H=0.001$ m and $\Omega =2$ rad/s.

**References**

- Abbott, J. R., N. Tetlow, A. L. Graham, S. A. Altobelli, E. Fukushima, L. A. Mondy, and T. S. Stephens, "Experimental observations of particle migration in concentrated suspensions: Couette flow," J. Rheol. 35, 773-795 (1991).
- Acrivos, A., "Shear-induced particle diffusion in concentrated suspensions of noncolloidal particles," J. Rheol. 39, 813-826 (1995).
- Aktas, S., D. M. Kalyon, B. M. Marín-Santibáñez and J. Pérez-González, "Shear viscosity and wall slip behavior of a viscoplastic hydrogel," J. Rheol. 58, 513-535 (2014).
- Allain, C., M. Cloitre, and M. Wafra, "Aggregation and sedimentation in colloidal suspensions," Phys. Rev. Lett. 74, 1478–1481 (1995).
- Allende, M., and D. Kalyon, "Assessment of Particle-Migration Effects in Pressure-Driven Viscometric Flows," J. Rheol. 44, 79-90 (2000).
- Altobelli, S. A., E. Fukushima, and L. A. Mondy, "Nuclear magnetic resonance imaging of particle migration in suspensions undergoing extrusion," J. Rheol. 41, 1105-1115 (1997).
- Aral, B., and D. M. Kalyon, "Effects of Temperature and Surface Roughness on Time-Dependent Development of Wall Slip in Torsional Flow of Concentrated Suspensions," J. Rheol. 38, 957-972 (1994).
- Aral, B., and D. M. Kalyon, "Viscoelastic material functions of noncolloidal suspensions with spherical particles," J. Rheol. 41, 599-620 (1997).
- Aral, B., and D. M. Kalyon, "Rheology and Extrudability of Very Concentrated Suspensions: Effects of Vacuum Imposition," Plast. Rubber and Comp. Proc. and Appl. 4, 201-210 (1995).
- Arola, D. F., G. A. Barrall, R. L. Powell, K. L. McCarthy, and M. J. McCarthy, "Use of nuclear magnetic resonance imaging as a viscometer for process monitoring," Chem. Eng. Sci. 52, 2049–2057 (1997).
- Baek, G., and C. Kim, "Rheological properties of Carbopol containing nanoparticles," J. Rheol. 55, 313-330 (2011).
- Ballesta, P., R. Besseling, L. Isa, G. Petekidis, and W. C. K. Poon, "Slip and flow of hard-sphere colloidal glasses," Phys. Rev. Lett. 101, 258301 (2008).
- Barnes, H. A., and J. O. Carnali, "The vane-in-cup as a novel rheometer geometry for shear thinning and thixotropic materials," J. Rheol. 36, 841–866 (1990).
- Baudez, J., S. Rodts, X. Chateau, and P. Coussot, "New technique for reconstructing instantaneous velocity profiles from vis- cometric tests: Application to pasty materials," J. Rheol. 48, 69–82 (2004).
- Bergenholtz, J., W. C. Poon, and M. Fuchs, "Gelation in model colloid-polymer mixtures," Langmuir 19, 4493–4503 (2003).
- Bird, B., R. Armstrong, and O. Hassager, "Dynamics of Polymeric Liquids," Volume 1, 2nd edition, John Wiley & Sons, NY (1987).
- Bird, R. B., W. E. Stewart, and E. N. Lightfoot, "Transport Phenomena," John Wiley & Sons Inc. New York (1960).
- Birinci, E., and D. M. Kalyon, "Development of extrudate distortions in poly(dimethyl siloxane) and its suspensions with rigid particles," J. Rheol. 50, 313-326 (2006).
- Boger, D. V., "Rheology of slurries and environmental impacts in the mining industry," Annu. Rev. Chem. Biomol. Eng. 4, 239-257 (2013).
- Bonn, D., and M. M. Denn, "Yield stress fluids slowly yield to analysis," Science, 324, 1401–1402 (2009).

- Bonn, D., S. Rodts, M. Groenink, S. Rafai, N. Shahidzadeh-Bonn, and P. Coussot, "Some applications of magnetic resonance imaging in fluid mechanics: Complex flows and complex fluids," Annu. Rev. Fluid Mech. 40, 209–233 (2008).
- Brunn, P. O., and J. Vorwerk, "Determination of the steady-state shear viscosity from measurements of the apparent viscosity for some common types of viscometers," Rheologica acta, 32, 380-397 (1993).
- Budtova, T. V., V. P. Budtov, P. Navard, and S. Y. Frenkel, "Rheological properties of highly swollen hydrogel suspensions," J. Appl. Polym. Sci. 52, 721–726 (1994).
- Buzzaccaro, S., R. Rusconi, and R. Piazza, ""Sticky" hard spheres: Equation of state, phase diagram, and metastable gels," Phys. Rev. Lett. 99, 098301 (2007).
- Cardinaux, F., T. Gibaud, A. Stradner, and P. Schurtenberger, "Interplay between spinodal decomposition and glass formation in proteins exhibiting short-range attractions," Phys. Rev. Lett. 99, 118301 (2007).
- Chen, Y., D. M. Kalyon and E. Bayramli, "Effects of Surface Roughness and the Chemical Structure of Materials of Construction on Wall Slip Behavior of Linear Low-Density Polyethylene in Capillary Flow," J. Appl. Polym. Sci. 50, 1169-1177 (1993).
- Cohen, Y., and A. B. Metzner, "Apparent slip flow of polymer solutions," J. Rheol. 29, 67–102 (1985).
- Coussot, P., L. Tocquer, C. Lanos, and G. Ovarlez, "Macroscopic vs. local rheology of yield stress fluids," J. Non-Newtonian Fluid Mech. 158, 85–90 (2009).
- De Rosa, M. E, and H. H. Winter, "The effect of entanglements on the rheological behavior of polybutadiene critical gels," Rheologica acta, 33, 220-237 (1994).
- Derakhshandeh, B., S. G. Hatzikiriakos, and C. P. J. Bennington, "Rheology of pulp suspensions using ultrasonic Doppler velocimetry," Rheologica acta, 49, 1127–1140 (2010).
- Divoux, T., C. Barentin, and S. Manneville, "From stress-induced flu- idization processes to Herschel-Bulkley behaviour in simple yield stress fluids," Soft Matter, 7, 8409–8418 (2011).
- Divoux, T., D. Tamarii, C. Barentin, S. Teitel, and S. Manneville, "Yielding dynamics of a Herschel-Bulkley fluid: A critical-like fluidi- zation behaviour," Soft Matter, 8, 4151–4164 (2012).
- El Kissi, N., and J. M. Piau, "The different capillary flow regimes of entangled polydimethlsiloxane polymers: Macroscopic slip at the wall, hysteresis and cork flow," J. Non-Newtonian Fluid Mech. 37, 55–94 (1990).
- Estellé, P., C. Lanos, A. Perrot, and S. Amziane, "Processing the vane shear flow data from Couette analogy," Appl. Rheol. 18, 34037–34481 (2008).
- Flory, P. J., "Principles of polymer chemistry", Cornell University Press (1953).
- Gauckler, L. J., T. Graule, and F. Baader, "Ceramic forming using enzyme catalyzed reactions," Mater. Chem. Phys. 61, 78–102 (1999).
- Gevgilili, H., and D. M. Kalyon, "Step Strain Flow: Wall Slip and other Error Sources", J. Rheol. 45, 2, 467-475 (2001).
- Gevgilili, H., D. Kalyon and A. Shah, "Processing of energetics in continuous shear roll mills," J. Energ. Mater. 26, 29-51 (2008).
- Giesekus H. and G. Langer, "Die bestimmung der wahren fliesskurven nicht-newtonischer flüssigkeiten und plastischer stoffe mit der methode der repräsentativen viscosität", Rheologica acta, 16, 1-22 (1977).
- Grant, M. C., and W. B. Russel, "Volume-fraction dependence of elastic moduli and transition temperatures for colloidal silica gels," Phys. Rev. E 47, 2606–2614 (1993).

- Graziano, R., V. Preziosi, D. Uva, G. Tomaiuolo, B. Mohebbi, J. Claussen, and S. Guido, "The microstructure of Carbopol in water under static and flow conditions and its effect on the yield stress," J. Colloid Interface Sci. 582, 1067-1074 (2021).
- He, J., S. Lee and D. Kalyon, "Shear viscosity and wall slip behavior of dense suspensions of polydisperse particles," J. Rheol. 63, 1, 19-32 (2019).
- Holenberg, Y., O. M. Lavrenteva, U. Shavit, and A. Nir, "Particle tracking velocimetry and particle image velocimetry study of the slow motion of rough and smooth solid spheres in a yield-stress fluid," Phys. Rev. E 86, 066301 (2012).
- Jana, S. C., B. Kapoor, and A. Acrivos, "Apparent wall slip velocity coefficients in concentrated suspensions of noncolloidal particles," J. Rheol. 39, 1123-1132 (1995).
- Jiang, T., A. Young, and A. B. Metzner, "The rheological characterization of HPG gels: Measurement of slip velocities in capillary tubes," Rheologica acta, 25, 397–404 (1986).
- Kalyon, D. M., R. Yazici, C. Jacob, B. Aral and S. W. Sinton, "Effects of Air Entrainment on the Rheology of Concentrated Suspensions during Continuous Processing," Polym. Eng. Sci. 31, 1386-1392 (1991a).
- Kalyon, D. M., C. Jacob and P. Yaras, "An Experimental Study of the Degree of Fill and Melt Densification in Fully-intermeshing, Co-rotating Twin Screw Extruders," Plastics, Rubber and Composites Processing and Applications, 16 (3), 193-200 (1991b).
- Kalyon, D. M., P. Yaras, B. Aral and U. Yilmazer, "Rheological Behavior of Concentrated Suspensions: A Solid Rocket Fuel Simulant," J. Rheol. 37, 35-53 (1993).
- Kalyon, D. M., "Review of Factors Affecting the Continuous Processing and Manufacturability of Highly Filled Suspensions," Journal of Materials Processing and Manufacturing Science, 2 159-187 (1993).
- Kalyon, D. M., H. Gokturk, P. Yaras and B. Aral, "Motion Analysis of Development of Wall Slip during Die Flow of Concentrated Suspensions," Society of Plastics Engineers ANTEC Technical Papers, 41, 1130-1134 (1995).
- Kalyon, D. M., "Highly Filled Materials: Understanding the Generic Behavior of Highly Filled Materials Leads to Manufacturing Gains and New Technologies," Chem. Tech. 25, 22-30 (1995).
- Kalyon, D. M., A. Lawal, R. Yazici, P. Yaras, and S. Railkar, "Mathematical modeling and experimental studies of twin-screw extrusion of filled polymers," Polym. Eng. Sci. 39, 1139-1151 (1999).
- Kalyon, D. M. and H. Gevgilili, "Wall slip and extrudate distortion of three polymer melts", J. Rheol. 47, 3, 683-699 (2003).
- Kalyon, D. M., E. Birinci and H. Gevgilili, "Development of extrudate distortions as affected by wall slip behavior of polymers and filled polymers," Proceedings of Annual Meeting of American Institute of Chemical Engineers (2003).
- Kalyon, D. M., "Comments on "A new method of processing capillary viscometry data in the presence of wall slip" [J. Rheol. 47, 337-348 (2003)]", J. Rheol. 47, 1087-1088 (2003).
- Kalyon, D. M., H. Gevgilili, and A. Shah, "Detachment of the polymer melt from the roll surface: data from a shear roll extruder," Int. Polym. Process. 19, 129-138 (2004).
- Kalyon, D. M., "Comments on the use of rheometers with rough surfaces or surfaces with protrusions," J. Rheol. 49, 1153-1155 (2005).
- Kalyon, D. M., "Apparent Slip and Viscoplasticity of Concentrated Suspensions", J. Rheol. 49, 621-640 (2005).

- Kalyon, D. M., D. Dalwadi, M. Erol, E. Birinci and C. Tsenoglou, "Rheological Behavior of Concentrated Suspensions as affected by the Dynamics of the Mixing Process", Rheologica acta, 45, 641-658 (2006a).
- Kalyon, D. M., H. Gevgilili, J. Kowalczyk, S. Prickett and C. Murphy, "Use of adjustable-gap on-line and off-line slit rheometers for the characterization of the wall slip and shear viscosity behavior of energetic formulations," J. Energ. Mater. 24, 175-193 (2006b).
- Kalyon, D. M., H. Tang and B. Karuv, "Squeeze flow rheometry for rheological characterization of energetic formulations," J. Energ. Mater. 24, 195-202 (2006c).
- Kalyon, D. M., and H. Tang, "Inverse problem solution of squeeze flow for parameters of generalized Newtonian fluid and wall slip," J. Non-Newtonian Fluid Mech. 143, 133 (2007).
- Kalyon, D. M., and M. Malik, "An integrated approach for numerical analysis of coupled flow and heat transfer in co-rotating twin screw extruders," Int. Polym. Process. 22, 293-302 (2007).
- Kalyon, D. M., "An Analytical Model for Steady Coextrusion of Viscoplastic Fluids in Thin Slit Dies with Wall Slip," Polym. Eng. Sci. 50, 652-664 (2010).
- Kalyon, D. M., and M. Malik, "Axial laminar flow of viscoplastic fluids in a concentric annulus subject to wall slip," Rheologica acta, 51, 805–820 (2012).
- Kalyon, D. M., "Yield stress and other flow and wall slip parameters of viscoplastic fluids from steady torsional flow," arXiv preprint arXiv:2106.13351 (2021).
- Ketz, R. J., R. K. Prud'homme, W. W. Graessley, "Rheology of concentrated microgel solutions," Rheologica acta, 27, 531-539 (1988).
- Koh, C. J., P. Hookham, and L. G. Leal, "An experimental investigation of concentrated suspension flows in a rectangular channel," J. Fluid Mech. 266, 1-32 (1994).
- Lawal, A., and D. M. Kalyon, "A Non-isothermal Model of Single Screw Extrusion Processing of Viscoplastic Materials Subject to Wall Slip," Society of Plastics Engineers ANTEC Technical Papers, 38, 2158-2161 (1992).
- Lawal, A., and D. M. Kalyon, "Extrusion of Viscoplastic Fluids Subject to Different Slip Coefficients at Screw and Barrel Surfaces," Society of Plastics Engineers ANTEC Technical Papers, 39, 2782-2785 (1993).
- Lawal, A., and D. M. Kalyon, "Non-isothermal Model of Single Screw Extrusion of Generalized Newtonian Fluids," Numer. Heat Transf. 26 (1), 103-121 (1994a).
- Lawal, A., and D. M. Kalyon, "Single Screw Extrusion of Viscoplastic Fluids Subject to Different Slip Coefficients at Screw and Barrel Surfaces," Polym. Eng. Sci. 34, 1471-1479 (1994b).
- Lawal, A., S. Railkar, P. Yaras and D. M. Kalyon, "Twin Screw Extrusion Processing of Filled Polymers," Society of Plastics Engineers ANTEC Technical Papers, 42, 381-385 (1996).
- Lawal, A., and D. M. Kalyon, "Viscous Heating in Nonisothermal Die Flows of Viscoplastic Fluids with Wall Slip," Chem. Eng. Sci. 52, 1323-1337 (1997a).
- Lawal, A., and D. M. Kalyon, "Non-Isothermal Extrusion Flow of Viscoplastic Fluids with Wall Slip," Int. J. Heat Mass Transf. 40, 3883-3897 (1997b).
- Lawal, A., and D. M. Kalyon, "Analysis of Nonisothermal Screw Extrusion Processing of Viscoplastic Fluids with Significant Backflow," Chem. Eng. Sci. 54, 999-1013 (1999).
- Lawal, A., and D. M. Kalyon, "Compressive Squeeze Flow of Viscoplastic Fluids with Apparent Wall Slip," Int. Polym. Proc. 15, 63-71 (2000).
- Leighton, D., and A. Acrivos, "The shear-induced migration of particles in concentrated suspensions," J. Fluid Mech. 181, 415-439 (1987).

- Lochhead, R. Y., "The role of polymers in cosmetics: recent trends," Cosmetic Nanotechnology, Chapter 1, pp 3-56 (2007).
- Lu, P. J., E. Zaccarelli, F. Ciulla, A. B. Schofield, F. Sciortino, and D. A. Weitz, "Gelation of particles with short-range attraction," Nature, 453, 499-503 (2008).
- Malik, M., and D. M. Kalyon, "Three-dimensional Finite Element Simulation of Processing of Generalized Newtonian Fluids in Counter-rotating and Tangential Twin Screw Extruder and Die Combination," Int. Polym. Proc. 20, 398- 409 (2005).
- Malik, M., D. M. Kalyon, and J. C. Golba Jr., "Simulation of co-rotating twin screw extrusion process subject to pressure-dependent wall slip at barrel and screw surfaces: 3D FEM Analysis for combinations of forward- and reverse-conveying screw elements," Int. Polym. Proc. 29, 51-62 (2014).
- Medina-Bañuelos, E. F., B. M. Marín-Santibáñez, J. Pérez-González, M. Malik and D. M. Kalyon, "Tangential annular (Couette) flow of a viscoplastic hydrogel with wall slip," J. Rheol. 61, 1007-1022 (2017).
- Medina-Bañuelos, E. F., B. M. Marín-Santibáñez, J. Pérez-González and D. M. Kalyon, "Rheo-PIV analysis of the vane in cup flow of a viscoplastic hydrogel," J. Rheol. 63, 905-915 (2019).
- Medina-Bañuelos, E. F., B. M. Marín-Santibáñez, and J. Pérez-González, "Rheo-PIV analysis of the steady torsional parallel-plate flow of a viscoplastic microgel with wall slip," J. Rheol. 66, 31-48 (2021).
- Meeker, S. P., R. T. Bonnecaze, and M. Cloitre, "Slip and flow in pastes of soft particles: Direct observation and rheology," J. Rheol. 48, 1295–1320 (2004a).
- Meeker, S. P., R. T. Bonnecaze, and M. Cloitre, "Slip and flow in soft particle pastes," Phys. Rev. Lett. 92, 198302 (2004b).
- Mooney, M., "Explicit formulas for slip and fluidity," J. Rheol. 2, 210–222 (1931).
- Moraczewski, T., H. Tang, and N. C. Shapley, "Flow of a concentrated suspension through an abrupt axisymmetric expansion measured by nuclear magnetic resonance imaging," J. Rheol. 49, 1409–1428 (2005).
- Nazari, B., R. H. Moghaddam, and D. Bousfield, "A three-dimensional model of a vane rheometer," Int. J. Heat Fluid Flow. 42, 289–295 (2013).
- Nott, P. R., and J. F. Brady., "Pressure-driven flow of suspensions: simulation and theory," J. Fluid Mech. 275, 157-199 (1994).
- Onogi, S, T. Matsumoto, and Y. Warashina, "Rheological properties of dispersions of spherical particles in polymer solutions," Trans. Soc. Rheol. 17, 175-90 (1973).
- Ortega-Avila, F., J. Pérez-González, B. M. Marín-Santibáñez, F. Rodríguez-González, S. Aktas, M. Malik and D. M. Kalyon, "Axial annular flow of a viscoplastic microgel with wall slip," J. Rheol. 60, 503-515 (2016).
- Ovarlez, G., F. Mahaut, F. Bertrand, and X. Chateau, "Flows and heterogeneities with a vane tool: MRI measurements," J. Rheol. 55, 197–223 (2011).
- Pérez-González, J., J. J. López-Durán, B. M. Marín-Santibáñez, and F. Rodríguez-González, "Rheo-PIV of a yield-stress fluid in a capillary with slip at the wall," Rheologica acta, 51, 937-946 (2012).
- Piau, J. M., "Carbopol gels: Elastoviscoplastic and slippery glasses made of individual swollen sponges: Meso-and macroscopic properties, constitutive equations and scaling laws," J. Non-Newton. Fluid Mech. 144, 1-29 (2007).
- Raynaud, J. S., P. Moucheront, F. Bertrand, J. P. Guilbaud, and P. Coussot, "Direct determination by nuclear magnetic resonance of the thixotropic and yielding behavior of suspensions," J. Rheol. 46, 709–732 (2002).

- Reiner, M., "Deformation, Strain and Flow, HK Lewis & Co" Ltd., London (1960).
- Roberts, G. P., and H. A. Barnes, "New measurements of the flow-curves for Carbopol dispersions without slip artefacts," Rheologica acta, 40(5): 499-503 (2001).
- Rodrigues, J. A., G. P. F. Oliveira, and C. M. Amaral, "Effect of thickener agents on dental enamel microhardness submitted to at-home bleaching," Brazilian oral research, 21, 170-175 (2007).
- Sahin, E., and D. Kalyon, "Preshearing is an in situ setting modification method for inorganic bone cements", Medical Devices and Sensors, 3, 2 (2020).
- Shafiei, M., M. Balhoff, and N. W. Hayman, "Chemical and microstructural controls on viscoplasticity in Carbopol hydrogel," Polymer, 139, 44-51 (2018).
- Shah, S. A., Y. L. Chen, K. S. Schweizer, and C. F. Zukoski, "Phase behavior and concentration fluctuations in suspensions of hard spheres and nearly ideal polymers," J. Chem. Phys. 118, 3350–3361 (2003).
- Sinton, S. W., and A. W. Chow, "NMR flow imaging of fluids and solid suspensions in Poiseuille flow," J. Rheol. 35, 735-772 (1991).
- Soltani, F., and U. Yilmazer, "Slip velocity and slip layer thickness in flow of concentrated suspensions," J. Appl. Polym. Sci. 70, 515–522 (1998).
- Sperling, L. H., "Introduction to Physical Polymer Science", 2nd Edition, John Wiley & Sons, New York, 357 (1992).
- Tanaka, H., Y. Nishikawa, and T. Koyama, "Network-forming phase separation of colloidal suspensions," J. Phys. Condens. Matter, 17, L143–L153 (2005).
- Tang, H. S., and D. M. Kalyon, "Estimation of the parameters of Herschel-Bulkley fluid under wall slip using a combination of capillary and squeeze flow viscometers," Rheologica acta, 43, 80-88 (2004a).
- Tang, H. S., and D. M. Kalyon, "Time-dependent tube flow of compressible suspensions subject to pressure dependent wall slip: Ramifications on development of flow instabilities," J. Rheol. 52, 1069-1090 (2008).
- Trappe, V., V. Prasad, L. Cipelletti, P. N. Segre, and D. A. Weitz, "Jamming phase diagram for attractive particles," Nature, 411, 772-775 (2001).
- Vand, V., "Viscosity of solutions and suspensions. I. Theory," J. Phys. Chem. 52, 277-299 (1948).
- Verduin, H., and J. K. Dhont, "Phase diagram of a model adhesive hard-sphere dispersion," J. Colloid Interface Sci. 172, 425–437 (1995).
- Vural, S., K. Dikovics and D. Kalyon, "Crosslink Density, Viscoelasticity and Swelling of Hydrogels as affected by Dispersion of Multi-walled Carbon Nanotubes", Soft Matter, 6, 3870-3875 (2010)
- Wang, Y., S. Aktas, S. A. Sukhishvili and D. M. Kalyon, "Rheological behavior and self-healing of hydrogen-bonded complexes of a triblock Pluronic® copolymer with a weak polyacid", J. Rheology 61, 6, 1103-1119 (2017).
- Winter, H. H., and F. Chambon, "Analysis of linear viscoelasticity of a crosslinking polymer at the gel point," J. Rheol. 30, 367-382 (1986).
- Yang, M. C., L. E. Scriven, and C. W. Macosko, "Some rheological measurements of magnetic iron oxide suspensions in silicone oil," J. Rheol. 30, 1015-29 (1986).
- Yaras, P., D. M. Kalyon, and U. Yilmazer, "Flow Instabilities in Capillary Flow of Concentrated Suspensions," Rheologica acta, 33, 48-59 (1994).

- Yilmazer, U., and D. M. Kalyon, "Slip effects in capillary and parallel disk torsional flows of highly filled suspensions," J. Rheol. 33, 1197-1212 (1989).
- Yilmazer, U., and D. M. Kalyon, "Dilatancy of Concentrated Suspensions with Newtonian Matrices," Polym. Compos. 12, 226-232 (1991).
- Yoshimura, A., and R. K. Prud'homme, "Wall slip corrections for Couette and parallel disk viscometers," J. Rheol. 32, 53-67(1988).